\documentclass[12pt, letterpaper]{article}
\pdfoutput=1
\usepackage[utf8]{inputenc}
\usepackage{graphicx}
\usepackage[margin=1in]{geometry} 
\usepackage{appendix} 
\usepackage{wrapfig} 
\usepackage[superscript,biblabel]{cite} 
\usepackage[skip=5pt,font=footnotesize,labelfont=bf]{caption}

\usepackage{authblk} 

\usepackage{amssymb,amsmath}
\usepackage{mathrsfs} 
\usepackage{booktabs} 
\usepackage{siunitx} 
\usepackage{textcomp}
\usepackage{xcolor} 

\graphicspath{ {images/} } 

\title{Pore-Scale Flow Characterization of Polymer Solutions\\in Microfluidic Porous Media}
\author[1]{Christopher A. Browne}
\author[1]{Audrey Shih}
\author[1,*]{Sujit S. Datta}
\affil[1]{Chemical and Biological Engineering, Princeton University, Princeton, NJ}
\affil[*]{Email: ssdatta@princeton.edu}

\date{9 September 2019} 

\begin{document}
\maketitle
\begin{abstract}
\noindent Polymer solutions are frequently used in enhanced oil recovery and groundwater remediation to improve the recovery of trapped non-aqueous fluids. However, applications are limited by an incomplete understanding of the flow in porous media. The tortuous pore structure imposes both shear and extension, which elongates polymers; moreover, the flow is often at large Weissenberg numbers Wi at which polymer elasticity in turn strongly alters the flow. This dynamic elongation can even produce flow instabilities with strong spatial and temporal fluctuations despite the low Reynolds number Re. Unfortunately, macroscopic approaches are limited in their ability to characterize the pore-scale flow. Thus, understanding how polymer conformations, flow dynamics, and pore geometry together determine these non-trivial flow patterns and impact macroscopic transport remains an outstanding challenge. Here, we describe how microfluidic tools can shed light on the physics underlying the flow of polymer solutions in porous media at high Wi and low Re. Specifically, microfluidic studies elucidate how steady and unsteady flow behavior depends on pore geometry and solution properties, and how polymer-induced effects impact non-aqueous fluid recovery. This work provides new insights for polymer dynamics, non-Newtonian fluid mechanics, and applications such as enhanced oil recovery and groundwater remediation.
\end{abstract}

\newpage
\section{Motivation}\label{sec:motivation}
Dissolving high molecular weight polymers in water increases the fluid viscosity and, for sufficiently concentrated solutions, imparts new elastic properties.  These effects can dramatically alter flow behavior, with important consequences in biological systems (e.g. pathogen clearance by mucus in the lungs), industrial applications (e.g. spinning and casting materials), and microfluidic devices (e.g. mixing for lab-on-a-chip applications). They are also frequently harnessed to improve enhanced oil recovery (EOR)\cite{sorbie2013} and groundwater remediation\cite{roote1998}. In these cases, an injected polymer solution must navigate a disordered porous medium, such as a reservoir rock or a subsurface aquifer, and displace trapped non-aqueous fluids such as oil or contaminants from the pore space, enabling them to be recovered downstream. In addition to approaches such as particle, surfactant, and foam injection,\cite{liu2007surfactant, yekeen2017,samanta2012, gbadamosi2019, raffa2016} polymer injection is emerging as an attractive approach to EOR and groundwater remediation for three key reasons: (i) Polymers can yield increased fluid recovery compared to water, (ii) Polymers are often more affordable than other additives such as polymer nanoparticles and specialized surfactants, and (iii) Because polymers impart new rheological properties to the fluid, they can give rise to new flow behaviors that can be combined with other techniques such as particle or surfactant injection. Being able to predict and control how macroscopic process variables like fluid pressure and total recovery depend on polymer solution properties and injection conditions is therefore of key practical and economic importance.

These variables are typically characterized at the macroscopic scale through measurements of pressure drop and fluid flow rate in sand packs, core samples, or entire formations. These measurements are then interpreted via Darcy's law: $|\Delta P|=\eta_{\text{app}}(Q/A)L/k$, where $\Delta P$ is the pressure drop across a length $L$ of the porous medium, $Q$ is the volumetric flow rate through a cross-sectional area $A$, $k$ is the medium permeability, which depends on the pore space geometry and amount of fluid trapping, and $\eta_{\text{app}}$ is the `apparent viscosity' of the polymer solution. Such studies provide important empirical measurements of $\eta_{\text{app}}$, which reflects the combined influence of different pore-scale flow behaviors. Intriguingly, $\eta_{\text{app}}$ is often found to increase abruptly above a threshold flow rate.\cite{james1975} This increase in flow resistance has in turn been linked to increased recovery of trapped non-aqueous fluid,\cite{wang2011,sandiford1964, pitts1995, wei2014, vermolen2014,clarke2016} challenging conventional belief that polymers cannot appreciably improve recovery.\cite{sorbie2013,du2004} However, despite its strong impact on EOR and groundwater remediation processes, the reason for this increased flow resistance and the concomitant increase in fluid recovery is still a puzzle. Many mechanisms have been proposed: resistance to polymer elongation,\cite{haward2003, odell2006} permeability reduction via polymer adsorption,\cite{zaitoun1988, zaitoun1998} suppressed breakup of the trapped fluid,\cite{huh2008} and the onset of unsteady flow fluctuations.\cite{pearson1976, larson1992,groisman2000,tsiklauri1994,clarke2016} In practice, multiple such mechanisms may be simultaneously at play. 

Unfortunately, macroscopic transport measurements do not provide access to the underlying flow behavior due to the opacity of typical 3D media. As a result, the conditions under which these mechanisms may arise are unknown. X-ray scanning and nuclear magnetic resonance measurements can provide valuable information of 3D structure and macroscopic flow, but cannot resolve pore-scale flow details with sufficient spatial or temporal resolution.\cite{mitchell2016, blunt2017, seymour1997, li2017, brown2017, shattuck1997, kutsovsky1996, lebon1996, lebon1997, sederman1997, sederman2001} Flow imaging in mesoscale model systems provides some key insights into the flow structure,\cite{ boger1987, lipscomb1988, mongruel1995, mongruel2003} but such devices cannot recapitulate the small length scales that characterize real-world porous media (Table \ref{tab:dimensions}). As a result, these
approaches are often subject to inertial effects, which typically do not arise during EOR and groundwater remediation and instead complicate the interpretation of laboratory measurements. As a result, the mechanisms by which polymer solutions potentially increase flow resistance and enhance fluid recovery remain poorly understood and hotly debated.\cite{zhu2012, urbissinova2010, zhang2011}

In this Review, we describe how microfluidic tools can shed light on this puzzle. First, we summarize how microfluidics provides a straightforward means to directly visualize flow behavior in model porous media (\S\ref{sec:microfluidic-models}). Next, we summarize the essential flow properties of bulk polymer solutions (\S\ref{sec:steady-bulk}). We then review the body of work using microfluidics to study how polymer dynamics and porous medium geometry together determine steady (\S\ref{sec:steady-extension}) and unsteady (\S\ref{sec:unsteady-extension}--\ref{sec:unsteady}) flow behavior. We then describe recent efforts applying these insights to understand the mechanisms underlying enhanced fluid recovery due to polymer solutions (\S\ref{sec:multiphase}). Finally, we highlight new opportunities for microfluidics in elucidating polymer solution flow in porous media, with consequences for EOR, groundwater remediation, and other emerging applications (\S\ref{sec:outlook}).

\section{Microfluidic models of porous media}\label{sec:microfluidic-models}
\begin{figure}[h]
	\centering
	\includegraphics[width=\textwidth]{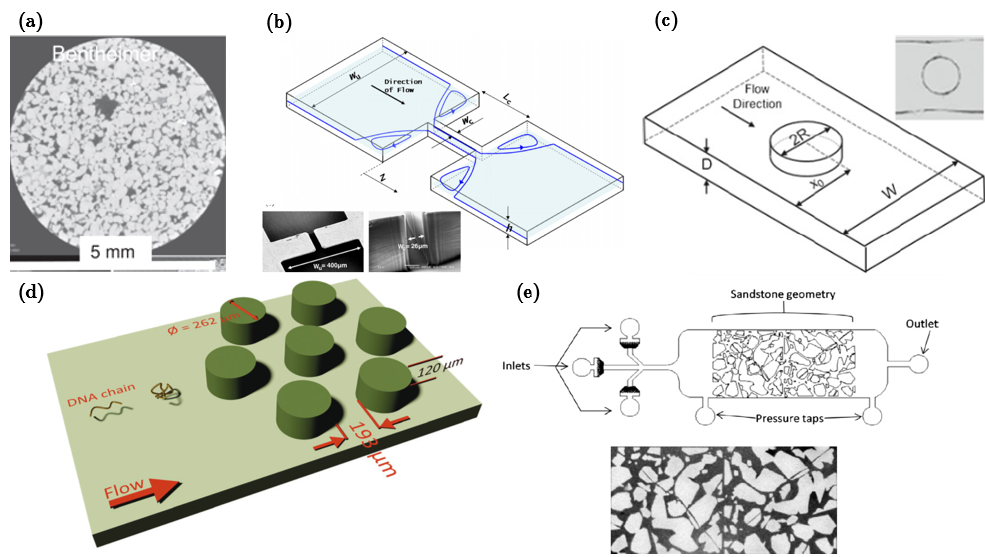}
	\caption{\textbf{(a)} X-ray CT scan of Bentheimer sandstone, a typical rock encountered in EOR,\cite{blunt2013} for comparison to several PDMS microfluidic model porous media. Copyright 2013, Elsevier; reprinted with permission. \textbf{(b)} Sudden flow constriction models the expansion-contracting geometry of the pore space.\cite{rodd2007} Copyright 2007, Elsevier; reprinted with permission. \textbf{(c)} A channel-centered cylinder models the impinging flow on a curved grain; $W$ is the channel width, $R$ is the cylinder radius, $x_0$ is the cylinder position, and $D$ is the channel height. Inset shows an photograph of a channel molded from a photolithographic cast.\cite{kenney2013} Copyright 2013, Springer; reprinted with permission. \textbf{(d)} A 2D array of cylindrical pillars to model a porous medium.\cite{kawale2017b} Copyright 2017, Royal Society of Chemistry; reprinted with permission. \textbf{(e)} A microfluidic device with a disordered 2D pore space, for more realistic modeling of a sandstone network.\cite{nilsson2013} Copyright 2013, Elsevier; reprinted with permission.}
	\label{fig:micromodels}
\end{figure}

The porous media relevant to EOR and groundwater remediation are typically dense packings of water-wet soil, rock, and mineral grains. The resulting pore spaces are highly heterogeneous and tortuous, with successive contractions and expansions known as pore throats and pore bodies, respectively; the pore throat diameters $L_\text{t}$ typically range from $\sim100$ nm to 10 $\mu$m (Figure \ref{fig:micromodels}a) while the pore body diameters $L_\text{b}$ are typically $\approx2$ to 5 times larger. During injection, a polymer solution is forced to flow through this pore space. The interstitial injection speed $U\equiv Q/\phi A$, where $\phi\approx0.1$ to 0.4 is the medium porosity, is limited by the ability to apply a sufficiently large fluid pressure drop; thus, interstitial flow speeds typically range between $\sim0.01$ and 100 $\mu$m/s.\cite{muggeridge2014, clarke2016} We summarize these pore-scale parameters in Table 1. Under these conditions, inertia is always absent: the fluid \textit{Reynolds number} $\mathrm{Re}\equiv\rho U L_\text{t}/\eta$, which describes the relative importance of inertial to viscous stress for a fluid with density $\rho$ and shear viscosity $\eta$, ranges from $\sim10^{-11}$ to $10^{-3}\ll1$ (Table 2). Thus, Newtonian fluids exhibit laminar, steady-state flow. By contrast, polymer solutions can exhibit dramatic flow instabilities in this range of $\mathrm{Re}$ due to the influence of the fluid elasticity, as we describe further in \S\ref{sec:unsteady}. 

Microfluidics provides the ability to design 	pore space geometries at length scales relevant to real-world porous media, with spatial resolutions as small as $\sim50$ nm. Furthermore, photolithography allows for fabrication of a wide range of complex microfluidic designs that can recapitulate pore space geometries, and can be used to systematically test the influence of different geometric parameters on flow behavior.\cite{anbari2018} Finally, microfluidic devices can be designed to be optically transparent, enabling pore-scale flow visualization using optical imaging. Such platforms thus enable systematic characterization of the flow patterns, polymer elongation dynamics, and flow fluctuations during polymer solution flow through model porous media. 2D microfluidic channels are typically fabricated in PDMS (polydimethylsiloxane) using soft photolithography, or in hard polymeric resins using 3D-printing.\cite{oconnell2019} These channels are then sealed against a glass slide to allow imaging of the flow using tracers via optical microscopy.\cite{thielicke2014}. Designs of varying complexity can then be used to either simplify or closely mimic the complex geometry of real-world porous media (Figure \ref{fig:micromodels}). 

\begin{table}[h!]
  \begin{center}
    \caption{Table of key descriptors of polymer solution flow in real-world porous media.}
    \label{tab:dimensions}
    \begin{tabular}{ll}
      \toprule 
       \textbf{Parameter} & \textbf{Order-of-magnitude range}\\
      \midrule 
            Pore throat diameter $L_\text{t}$ & 0.1 to $10$ $\mu$m\\
            Pore body diameter $L_\text{b}$ & 0.2 to $50$ $\mu$m\\
            Fluid viscosity $\eta$ & 1 to 100 cP \\
            Interstitial flow speed $U$& $10^{-2}$ to $10^{2}$ $\mu$m/s\\
            Shear rate $\dot{\gamma}\sim U/L_\text{t}$ & $10^{-3}$ to $10^{3}$ $\text{s}^{-1}$\\
            Pore residence time $\tau_\text{res}\sim L_\text{b}/U$& $2\times10^{-3}$ to $5\times10^{3}$ s\\
            Polymer relaxation time $\lambda$& $10^{-2}$ to $10^{2}$ s\\
      \bottomrule 
    \end{tabular}
  \end{center}
\end{table}

\begin{table}[h!]
  \begin{center}
    \caption{Table of key dimensionless parameters characterizing polymer solution flow in porous media. $\rho$ is the fluid density, $U$ is the interstitial flow speed, $L_{t}$ is the pore throat diameter, $\eta$ is the fluid shear viscosity, $N_1$ is the first normal stress difference, $\sigma$ is the  shear stress, $\lambda$ is the polymer relaxation time, $\dot{\gamma}$ is the shear rate, $\tau_\text{res}$ is the polymer residence time in a pore, $\phi$ is the medium porosity, and $\gamma$ is the aqueous-nonaqueous fluid interfacial tension.}
    \label{tab:table1}
    \begin{tabular}{l l l l}
      \toprule 
       \textbf{Parameter} & \textbf{Definition} & \textbf{Interpretation} & \textbf{Range}\\
      \midrule 
     Reynolds &  $\text{Re}\equiv\rho U L_\text{t}/\eta$ & \footnotesize{inertial stress / viscous stress} & $10^{-11}$ to $10^{-3}$\\
            Weissenberg &  $\text{Wi}\equiv N_1/2\sigma\approx\lambda\dot{\gamma}$ & \footnotesize{elastic stress / viscous stress} & $10^{-5}$ to $10^{5}$\\
     Deborah &  $\text{De}\equiv\lambda/\tau_\text{res}$ & \footnotesize{polymer relaxation time / residence time} & $2\times10^{-6}$ to $5\times10^{4}$\\ 
     Capillary & $\text{Ca}\equiv\eta U\phi/\gamma$ & \footnotesize{pore-scale viscous stress / capillary pressure} & $10^{-10}$ to $10^{-3}$\\
      \bottomrule 
    \end{tabular}
  \end{center}
\end{table}

\begin{wrapfigure}{R}{0.47\textwidth}
	\begin{centering}
		\includegraphics[width=3.15in]{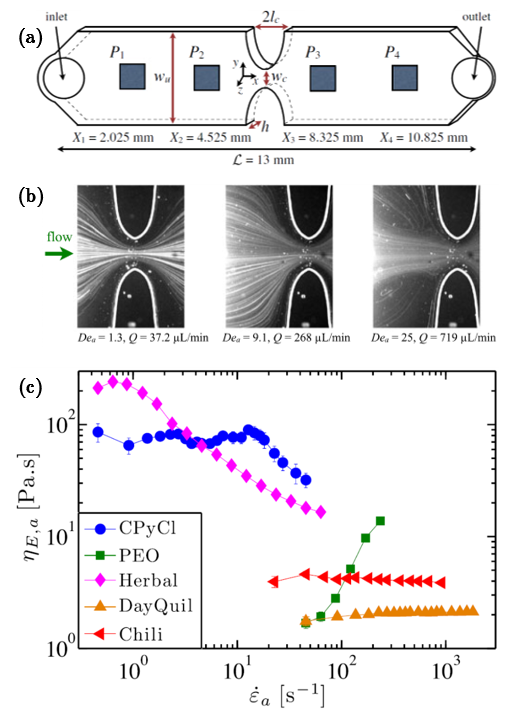}
	\end{centering}
\caption{\textbf{(a)} Microfluidic channel with hyperbolic constriction for extensional rheometry. The expansion and contraction widths are $w_u=2920$ $\mu$m and $w_c=400$ $\mu$m. \textbf{(b)} Streakline images of flow for 3000 ppm of 2 MDa PEO (a flexible polymer) using seed tracer particles; at the highest flow rate large eddies form in the corners leading into the constriction. \textbf{(c)} Extensional rheology for several different solutions measured with hyperbolic microchannel. CPyCl is a worm-like micelle solution composed of 100 mM cetylpyridinium
chloride and 60 mM sodium salicylate; PEO is a flexible polymer; Herbal is a commercially available surfactant solution of SDS-SLES; DayQuil is a commercially available solution of CMC, a rigid polymer; Chili is a commercially available solution of xanthan, a semi-rigid polymer.\cite{ober2013} Copyright 2013, Springer; reprinted with permission.}	\label{fig:hyperbola}
\end{wrapfigure}

The simplest representation of a pore is a sudden rectangular constriction, which captures the  expansion and contraction of the tortuous pore space (Figure \ref{fig:micromodels}b). Alternatively, the flow can be treated as impinging directly onto a curved grain, which can be represented as a channel-centered cylinder (Figure \ref{fig:micromodels}c). Both the rectangular constriction and channel-centered cylinder can be repeated in a linear series to model flow through multiple successive pores. More complex designs probe the spatial effects in a porous network by placing obstacles (square, circular, or triangular) on a 2D lattice (Figure \ref{fig:micromodels}d). At an even higher level of complexity, 2D sections of real porous media can be used to fabricate disordered porous networks (Figure \ref{fig:micromodels}e). While all of these designs are two-dimensional, and hence underrepresent the connectivity and complexity of the full three-dimensional pore space of real-world media, they provide valuable insights into polymer solution flow---specifically, how it is influenced by flow conditions and the geometry of the pore space.

\section{Steady Flow Properties of Bulk Polymer Solutions}\label{sec:steady-bulk}
\subsection{Shear and Extensional Viscosity}\label{sec:viscosity}
EOR and groundwater remediation efforts increasingly utilize high molecular weight (typically $\sim\textrm{MDa}$) water-soluble polymers, such as xanthan gum and polyacrylamide derivatives. In practice, these polymers are typically added to water at concentrations $C\sim$ 10 to 1000 ppm. These concentrations range from the dilute regime---in which the different polymer chains are well-separated and do not overlap---to the semi-dilute regime in which neighboring chains begin to interact; the transition between these two regimes is given by the ``overlap concentration'' $C^*$. 

One standard method of characterizing the solution flow behavior is using a shear rheometer, which quantifies the dependence between the shear stress $\sigma$ and the shear rate $\dot{\gamma}$: $\sigma=\eta_\text{s}(\dot{\gamma})\dot{\gamma}$, where the non-Newtonian shear viscosity $\eta_\text{s}$ is shear rate-dependent. In particular, these solutions are typically shear thinning. The addition of polymers monotonically increases the zero-shear viscosity of the solution; as shear rate increases, the viscosity decreases, ultimately approaching the background solvent viscosity. This shear rate-dependence is often modeled as a power law for ease, or with the Carreau-Yasuda model, which provides an analytical fit that is often used in numerics and simulations.\cite{graessley2008,balhoff2006} However, when the polymer solution is highly dilute, or the background solvent is highly viscous, shear-thinning is typically ignored. The solution---termed a Boger fluid---then has a nearly constant shear viscosity over a broad range of shear rates. Interestingly, some simulations suggest that shear-thinning effects are often unimportant in polymer solution flow through a porous medium, and flow behaviors are primarily determined by the coupling between pore space geometry and the elastic properties of the fluid.\cite{zami2016} Others, however, contend that shear-thinning effects dominate the flow behavior.\cite{machado2016}

Extensional flows also arise as the fluid traverses successive expansions and contractions in the pore space. This form of flow produces an additional deviation from the Newtonian viscosity in the form of an extensional viscosity, which is absent in Newtonian fluids under typical conditions: $\eta_\text{e}=\sigma_\text{e}/\dot{\varepsilon}$, where $\sigma_\text{e}$ is the extensional stress and $\dot{\varepsilon}$ is the extension rate. For rigid polymers, this extensional viscosity depends on concentration and polymer aspect ratio, the longest dimension of the rod-shaped polymer divided by its dimension across.\cite{batchelor1971} For polyelectrolytes, the extensional viscosity can also depend sensitively on salt concentration, the valency of the salt ions, and the particular cations and anions that associate with the chain.\cite{dobrynin1995, turkoz2018, kawale2017a} Measuring these effects can be challenging, however: most rheometers inherently have shear, making it difficult to differentiate the role of non-Newtonian shear and extensional viscosity. Extensional viscosity can be estimated using opposing jet flow, but it is again difficult to remove the effects of shear. Furthermore, these experiments inherently suffer from inertial effects due to their macroscopic scale.\cite{fuller1987, cathey1988} In some cases, the extensional viscosity can be estimated by tracking the pinch-off dynamics of thin fluid filaments, using a model that describes the balance between capillary and viscous effects.\cite{jimenez2018, walter2019}

By enabling the design of channels that generate flows with pure extension, microfluidics provides a valuable platform for accurate measurements of extensional viscosity.\cite{del2017} For example, stagnation point flows can provide pure planar extension with no shear or inertial effects, allowing direct measurement of the extensional viscosity.\cite{haward2016microfluidic} Channels with hyperbolic constrictions can also be designed to minimize shear effects;\cite{ober2013} pressure drop measurements across the constriction, corrected for shear contributions, then provide a direct measure of the extensional viscosity (Figure \ref{fig:hyperbola}). Other microfluidic designs use the breakup of aqueous droplets in an oil sheath fluid; the extensional viscosity of the polymer solution can then be obtained from the filament thinning dynamics.\cite{juarez2011, ingremeau2013, sachdev2016} Intriguingly, such microfluidic experiments have revealed that for sufficient extension rates, the extensional viscosity can abruptly increase---similar to transport measurements on bulk porous media\cite{kim2018}---suggesting a potential contribution to the increased macroscopic flow resistance.\\

\subsection{Quantifying flow-induced deformations}\label{sec:dimensionless}

For both shear-thinning and Boger polymer solutions, elastic effects arise from the deformations of the individual polymer chains. For example, shear and extensional flows can force the chains to elongate; this elongation then relaxes over a time scale $\lambda$. In certain cases this relaxation can be slow, leading to ``memory" of past deformation in the form of retained strain. These fluids are therefore viscoelastic, and exhibit many unusual phenomena that have been studied extensively: rod climbing, die swell, elastic recoil, tubeless siphoning, pressure hole errors, enhanced vortices at contractions, negative wakes, reduction of turbulent drag, and modified jet breakup.\cite{bird1987} 

This strain retention can be described using constitutive laws that track the accumulation, dissipation, and advection of polymer strain under flow; prominent examples include the Upper Convected Maxwell, Oldroyd-B, and FENE-P models.\cite{bird1987} The Upper Convected Maxwell model gives the fluid stress as a linear combination of a Hookean elastic solid and a Newtonian viscous fluid, with the added consideration that polymer strain is advected by the fluid flow. The Oldroyd-B model gives a more sophisticated treatment of polymers as infinitely extensible linear springs, which couple to the bulk fluid through a relaxation time and a retardation time. The FENE-P model (Finite-Extensibility non-linear Elastic model developed by Peterlin) resolves the non-physical feature of infinite chain extensibility by incorporating a non-linear diverging force needed to extend a polymer to its full contour length.\cite{peterlin1961}

Though tracking polymer strain can be computationally challenging, these physical effects can also be captured by two dimensionless parameters that quantify the characteristic amount of chain stretching in a given flow. The \textit{Weissenberg number} indicates the relative magnitude of elastic stresses, quantified by the first normal stress difference $N_1$, to viscous stresses, quantified by the shear stress $\sigma$: $$\mathrm{Wi}\equiv\frac{N_1\left(\dot{\gamma}\right)}{2\sigma\left(\dot{\gamma}\right)}\approx\lambda\dot{\gamma}.$$ \noindent
The factor of 1/2 arises from the Oldroyd-B constitutive model; some researchers omit this factor for simplicity, to obtain an order-of-magnitude estimate. The normal stress difference $N_1$ arises from the stretching of polymer molecules under curved streamlines, which generates a radially inward tension; both $N_1$ and $\sigma$ thus depend on the shear rate $\dot{\gamma}$ and can be measured using bulk rheology. A large value of Wi indicates that elastic effects are important, and is generally associated with large amounts of chain extension. 

The approximate form on the right hand side of the above equation is Weissenberg's initially proposed definition,\cite{weissenberg1947} but is only used when the fluid rheology is unknown. In this case, $\lambda$ is an appropriately chosen relaxation time of the polymer, commonly chosen to be the longest Rouse time, which describes the diffusion of the individual polymer segments but neglects hydrodynamic interactions between them.\cite{howe2015} However, other experimental measures can be used as well, such as the Zimm relaxation time that also incorporates hydrodynamic interactions between segments.\cite{macosko1994} This timescale is compared with the shear deformation time scale $\dot{\gamma}^{-1}$; however, this value can be replaced by the extensional time scale $\dot{\varepsilon}^{-1}$ in the case of extensional flow, giving the extensional $\mathrm{Wi}_\text{e}\equiv\lambda \dot{\varepsilon}$. 

Polymers used in EOR and groundwater remediation have $\lambda$ ranging from $\sim10^{-2}$ to $10^{2}$ s;\cite{clarke2016} moreover, estimating the characteristic deformation rate as the pore flow speed divided by the pore dimension yields a shear rate ranging from $\sim10^{-3}$ to $10^{3}$ $\text{s}^{-1}$. Thus, flow in a porous medium can be characterized by a broad range of Wi between $\sim 10^{-5}$ and $10^{5}$ (Table \ref{tab:table1}), indicating that in many cases shear and extensional forces due to flow through the tortuous pore space can extend polymer chains. Elastic effects can therefore play a considerable role.

Another common dimensionless parameter is the \textit{Deborah number}. Though often used interchangeably with Wi, De compares the polymer relaxation time to a flow residence time scale $\tau_\text{res}$, and not the deformation time scale:\cite{dealy2010, poole2012}$$\mathrm{De}\equiv\frac{\lambda}{\tau_\text{res}}.$$ \noindent Thus, a large value of De indicates that the polymer chains are being strained over timescales much shorter than their relaxation time---hence, the flow drives them away from their equilibrium state. Indeed, exploring the rheology of polymer solutions in the regime of intermediate and large Deborah numbers is an active area of research.\cite{hyun2011, rogers2018} In laboratory extension experiments (\S\ref{sec:unsteady-extension}) this flow residence time represents the duration over which a stress is applied. In a tortuous pore space, this residence time is given by the transit time between successive pore throat constrictions $\tau_\text{res}\sim L_\text{b}/U$; as a result, De can range between $\sim10^{-6}$ and $10^{5}$ (Table 2).

\section{Steady Flow of Polymer Solutions in Porous Media} \label{sec:steady-porous}
\subsection{Extensional Effects in Bulk Porous Media}\label{sec:steady-extension}
Most polymer solutions applied in EOR and groundwater remediation are shear-thinning or Boger fluids; thus, extensional effects resulting from flow through the expanding and contracting pore space are thought to be the primary cause of the increased macroscopic flow resistance observed in bulk porous media.\cite{cheng2010} This increased macroscopic flow resistance can arise from extensional flow in several different ways. 

First, for the case of semi-rigid and rigid polymers---such as xanthan, carboxymethylcellulose (CMC), and polyimide---flow imaging in mesoscale setups has revealed the the generation of large corner eddies upstream and downstream of sudden constrictions.\cite{lipscomb1988, boger1987}.  Eddies are caused by the preferential alignment of polymers entering the constriction to minimize the extensional stress associated with chain misalignment.\cite{batchelor1971,mongruel1995, mongruel2003} Given these results, the formation of eddies---which pose an additional source of dissipation and thus contribute to the macroscopic pressure drop\cite{mongruel2003}---may be expected in the many expansions and contractions of a real 3D porous medium. However, this hypothesis remains to be explored.

\begin{figure}[h]
	\centering
	\includegraphics[width=\textwidth]{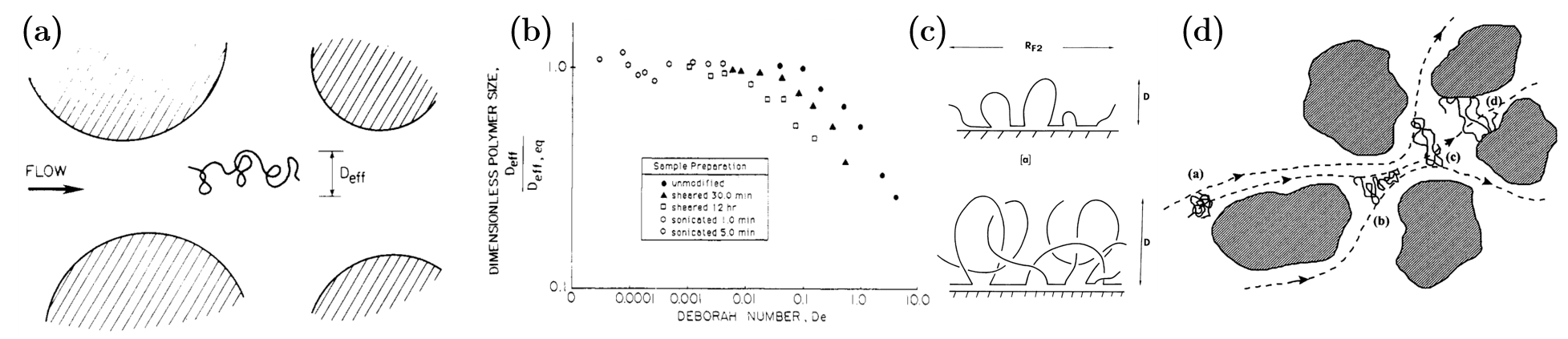}
\caption{\textbf{(a)} The extensional flow in porous media elongates flexible polymers.\cite{hoagland1989} Copyright 1989, American Chemical Society; reprinted with permission. \textbf{(b)} This effect can be measured in chromatagraphic columns because of the reduced effective size $\mathrm{D}_{\mathrm{eff}}$ which reduces the column transit time; higher $\text{Wi}_\text{e}$ (termed De here) elongate polymers more, decreasing the effective size.\cite{hoagland1989} Copyright 1989, American Chemical Society; reprinted with permission.  \textbf{(c)} Elongated chains can adsorb onto pore walls because of the increased number of contact points; hanging tails and loops can further entangle polymers at the wall.\cite{deGennes1976} Copyright 1976, EDP Sciences; reprinted with permission. \textbf{(d)} In low-permeability pores, extended polymers can bridge pore throats during adsorption, leading to clogging and flow redirection.\cite{zitha2001} Copyright 2001, Elsevier; reprinted with permission. }
	\label{fig:hoagland}
\end{figure}

Second, for the case of flexible polymers---such as polyacrylamide derivatives, poly(ethylene oxide), and DNA---extensional flows can stretch the individual chains in addition to orienting them (Figure \ref{fig:hoagland}a).\cite{hoagland1989, babcock2003} This change in polymer conformation can be indirectly deduced for flow through a porous medium using transit time measurements: the transit time of a polymer decreases as it elongates. Thus, measurements of polymer transit time for different imposed flow rates, corresponding to different characteristic extensional rates $\dot{\varepsilon}$, provide a measure of polymer elongation.\cite{hoagland1989} These measurements elegantly reveal that flexible polymers experience a coil-stretch transition, abruptly elongating above a threshold $\mathrm{Wi}_\text{e}\sim0.5$ (Figure \ref{fig:hoagland}b), in agreement with kinetic dumbbell extension theory and microfluidic studies of the coil-stretch transition\cite{schroeder2003} as detailed further in \S\ref{sec:unsteady-extension}. Polymers take a finite time to reach these equilibrium extensions, which can lead to transient effects as discussed further in \S\ref{sec:unsteady} for fast flows of high-molecular weight polymers in tortuous porous media. 

This flow-induced elongation can have important consequences for pore-scale and macroscopic transport. For example, as chains elongate, they can adsorb irreversibly onto pore walls, which in some cases leads to transient flow redirection or clogging. Coiled polymers typically have small binding energies, but the higher contact area of an elongated polymer produces strong, irreversible binding (Figure \ref{fig:hoagland}c).\cite{deGennes1976,obey1999, rubenstein2003} Adsorbed polymer layers can constrict pores and generate additional flow resistance (Figure \ref{fig:hoagland}d), potentially generating higher viscous pressure drops or redirecting the flow to bypassed regions, thereby promoting displacement and recovery of a trapped non-aqueous fluid.\cite{zaitoun1988, zaitoun1998,zitha1995, zitha2001,darwish2003} However, in high permeability reservoirs---where polymers have been successfully implemented---the typical thickness of adsorbed layers measured in experiments is too thin to dramatically alter flow resistance and fluid recovery.\cite{grattoni2004} This observation suggests that adsorption may play a key role, but cannot fully explain increased flow resistance or enhanced fluid recovery observed in bulk porous media. 

Another consequence of flow-induced elongation is an increase in the extensional viscosity, and hence flow resistance, due to the polymer chains resisting elongation.\cite{haward2003, odell2006} The onset of elongation in packed columns\cite{hoagland1989} coincides roughly with the critical shear rate for the excess resistance observed in model porous constrictions.\cite{chauveteau1981, chauveteau1984} Theoretical work\cite{prabhakar2004} was able to match this abrupt increase in extensional viscosity\cite{gupta2000} to the rapid coil-stretch transition\cite{smith1998}, indicating that polymer elongation is directly responsibly for this increased resistance. Fully capturing the physics of this process requires an understanding of the molecular details of polymer elongation; thus, microfluidic tools have been used to investigate the elongation dynamics of single chains. Indeed, simulations in porous media geometries exhibit similar increases in flow resistance, but only when using a polymer constitutive law that explicitly tracks extensional strain such as Oldroyd-B or FENE-P, highlighting the importance of polymer elongation.\cite{pilitsis1989, pilitsis1991} However, simulations applying these constitutive laws to complex porous geometries cannot fully reproduce the experimentally observed increase in flow resistance,\cite{talwar1995} suggesting that other dissipative processes arise in these flows, as we describe further in \S\ref{sec:unsteady}.

\subsection{Elongation Dynamics in Microfluidic Devices}\label{sec:unsteady-extension}

Microfluidic experiments provide a straightforward way to probe the elongation dynamics of polymers in extensional flows. These studies typically employ fluorescently-labeled DNA whose molecular weight can be tuned by harvesting it from different bacteriophage sources or by using restriction endonucleases. This capability allows for direct visualization of single polymer conformations.\cite{rye1993, teclemariam2007, hemminger2010, perkins1997, smith1998} For example, single molecules of DNA can be trapped at the stagnation point in a microfluidic cross-slot channel, enabling the dynamics of elongation to be monitored as the extensional flow is varied.\cite{rems2016, schroeder2018} Additionally, the DNA can be embedded within a background polymer solution, acting as a probe of the polymer conformations in the more concentrated solution.\cite{francois2008,francois2009}

In general, a polymer chain's equilibrium elongation increases monotonically with $\mathrm{Wi}_\text{e}$, transitioning gradually from a more coiled state at $\mathrm{Wi}_\text{e}=0$ to a more stretched state at higher $\mathrm{Wi}_\text{e}\gtrsim 0.5$. The chain length eventually plateaus as it approaches the contour length, which is the maximum elongation possible.\cite{bird1987} For longer chains, however, the transition from a coiled to a stretched state is sharp, yielding a nearly-discontinuous first-order transition.\cite{deGennes1974, fuller1980, hunkeler1996, menasveta1991, goff1994} In this case, the chain remains nearly coiled at low $\mathrm{Wi}_\text{e}$, showing only a slight increase in its equilibrium elongation as $\mathrm{Wi}_\text{e}$ is increased. Above a threshold $\mathrm{Wi}_\text{e}\sim0.5$, however, the equilibrium elongation increases sharply with $\mathrm{Wi}_\text{e}$. Further increases in $\mathrm{Wi}_\text{e}$ generate only a gradual increase in elongation as the chain approaches its contour length. This behavior allows a simple approximation of the conformation of large polymers: below $\mathrm{Wi}_\text{e}<0.5$ chains are fully coiled, and above $\mathrm{Wi}_\text{e}>0.5$ chains are fully stretched, known as the coil-stretch transition. 


\begin{figure}[h]
	\centering
		\includegraphics[width=4.5in]{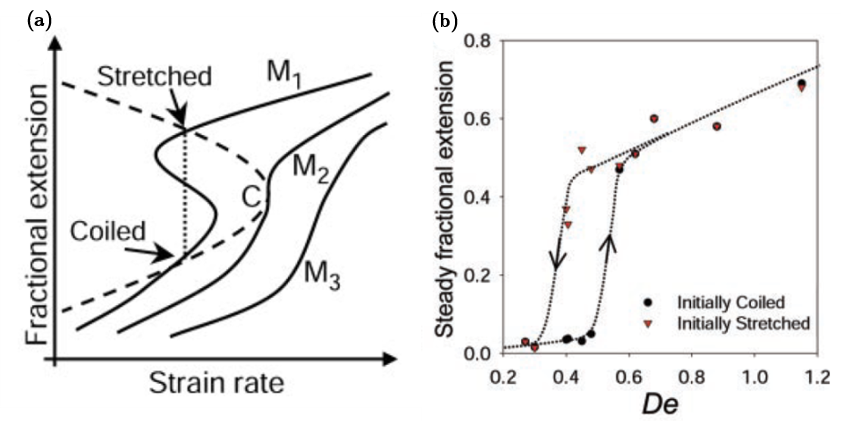}
\caption{\textbf{(a)} Sketch of hypothesized coil-stretch hysteresis for higher molecular weight polymers ($\mathrm{M}_1>\mathrm{M}_2>\mathrm{M}_3$), drawing an analogy to the liquid-vapor binodal in the style of de Gennes.\cite{schroeder2003, deGennes1974} \textbf{(b)} Equilibrium conformations of DNA from direct observation in a microfluidic cross-slot trap show a hysteresis window about $\mathrm{Wi}_\text{e}\sim0.4\pm0.1$ (termed De here), consistent with the hypothesized hysteresis window in \textbf{(a)}.\cite{schroeder2003} Copyright 2003, AAAS; reprinted with permission.}
	\label{fig:hysteresis}
\end{figure}

\begin{wrapfigure}{R}{0.5\textwidth}
	\begin{centering}
		\includegraphics[width=3.125in]{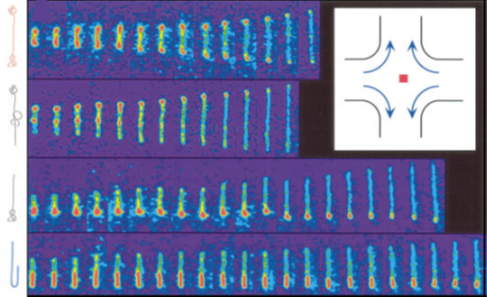}
	\end{centering}
\caption{Kinetically trapped conformations of DNA in an extensional trap after rapid startup flow show several different transient conformations based on initial conditions: dumbbell, kinked, half-dumbbell, and folded (top to bottom).\cite{perkins1997} Copyright 1997, AAAS; reprinted with permission.}
	\label{fig:kink}
\end{wrapfigure}

However, for large polymers in highly viscous solvents, the equilibrium length is not uniquely determined by $\mathrm{Wi}_\text{e}$---it also depends on the initial chain conformation, due to the difference in the hydrodynamic drag force exerted by the fluid on a coiled polymer compared to a stretched polymer.\cite{schroeder2003} Specifically, at intermediate $\text{Wi}_\text{e}\approx0.5$, the drag force on a coiled polymer is insufficient to extend it, but is large enough to maintain the extension of an initially stretched polymer. This difference in hydrodynamic drag creates a hysteretic dependence of polymer extension on strain history: with increasing $\mathrm{Wi}_\text{e}$, polymer stretching occurs at $\mathrm{Wi}_\text{e,stretch}\approx0.6$, while with decreasing $\mathrm{Wi}_\text{e}$, coiling occurs at a lower $\mathrm{Wi}_\text{e,coil}\approx0.4$ (Figure \ref{fig:hysteresis}). This hysteresis is a signature of a discontinuous first-order transition as theoretically predicted,\cite{deGennes1974} and thus both coiled and stretched polymers can coexist for $\mathrm{Wi}_\text{e}\approx0.5\pm0.1$. This bistability in conformations and coil-stretch hysteresis manifests in a broad range of conditions, including in the unstable flows described in \S\ref{sec:unsteady}.\cite{gerashchenko2005} 

In flows with intermediate to high Deborah numbers $\text{De}\gtrsim1$, the polymer chains do not have adequate time to reach these equilibrium conformations---instead, transient elongation and relaxation dynamics dominate.\cite{tanner1975, magda1988, rallison1988} Microfluidic measurements of transient single-polymer relaxation have provided scaling relationships for polymer relaxation and retardation times, which change in the dilute and semi-dilute regimes.\cite{liu2007, hsiao2017} Moreover, in startup flows, polymers can overshoot their equilibrium conformations before relaxing\cite{hur2001}---even becoming kinetically trapped in kinked or knotted conformations. These kinks and knots can strongly increase the rate of polymer elongation, and hence the `retardation time' needed by the chain to reach equilibrium.\cite{smith1998} Direct visualization of dilute DNA reveals four dominant conformations during transient elongation: dumbbell, kinked, half-dumbbell, and folded (Figure \ref{fig:kink}).\cite{perkins1997} For more concentrated semi-dilute solutions, interactions with other chains also influence the kinetics of stretching;\cite{hsiao2017} for example, kinks and folds are typically suppressed, and chains can instead have coiled ends, which again increase the polymer retardation time. Thus, for startup or highly time-dependent flows, kinetic trapping of polymers can lead to many more unique conformations, which may influence bulk flow properties.

\begin{figure}[h]
	\centering
		\includegraphics[width=\textwidth]{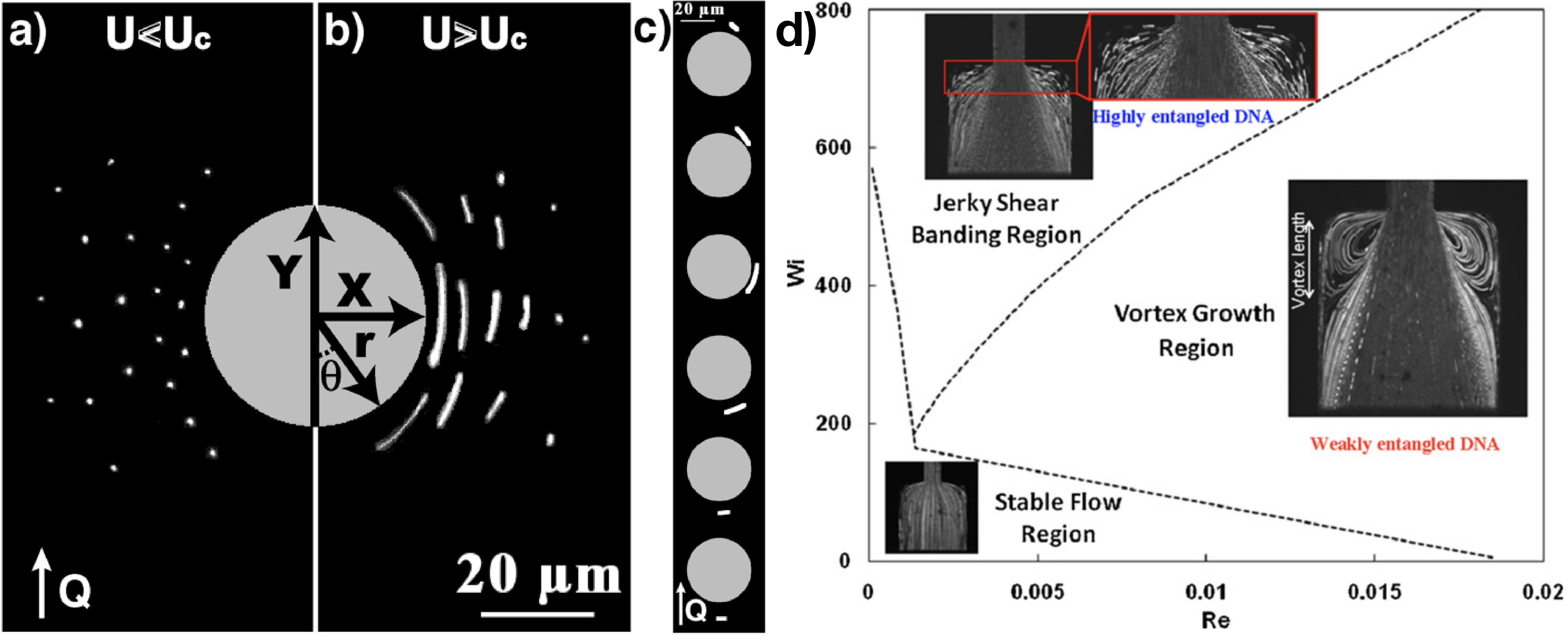}
	\caption{\textbf{(a-c)} Fluorescent micrographs of labeled DNA molecules flowing around a cylinder in a microfluidic channel (flow is bottom to top) at \textbf{(a)} a low flow rate below the coil-stretch transition, or at \textbf{(b)} a high flow rate above the coil-stretch transition. The chain elongation is strongest closest to the cylinder. \textbf{c)} A sequence of images showing the DNA molecule approaching the cylinder and elongating when it is sufficiently close.\cite{francois2008} Copyright 2008, American Physical Society; reprinted with permission. \textbf{(d)} Streakline images of fluorescent tracer particles shows the generation of large vortices at the entrance of a sudden contraction. Visualization of fluorescently labeled DNA confirms that DNA elongation drives the eddy growth. For highly entangled solutions, the eddies show transient fluctuations.\cite{hemminger2010} Copyright 2010, Elsevier; reprinted with permission.}
	\label{fig:DNAconstriction}
\end{figure}

Direct visualization using microfluidics confirms that polymer chains are abruptly elongated as they transit through a microfluidic pore space. Individual polymers flowing around channel-centered cylindrical obstacles are highly stretched (Figure \ref{fig:DNAconstriction}a-c), producing excess flow resistance consistent with the abrupt increase of $\eta_\text{app}$ in bulk porous media.\cite{francois2008} These elongated chains in turn align at a sudden contraction to form large vortices (Figure \ref{fig:DNAconstriction}d).\cite{hemminger2010} The mechanism for eddy formation is similar to the case of rigid polymers (\S\ref{sec:steady-extension}): eddy formation minimizes the extensional stress penalty of misaligned chains. 

Notably, these polymer chain conformations are highly time-dependent, leading to transient flow effects.\cite{tanner1975, magda1988, rallison1988} For example, in the simple case of a sudden constriction, the dynamic stretching and coiling of highly entangled polymers can produce time-dependent shear-banding and eddies with fluctuating surfaces (Figures \ref{fig:hyperbola} and \ref{fig:DNAconstriction}d).\cite{ober2013, hemminger2010} In micromodels with a channel-centered cylinder, transient polymer extension can lead to a time-dependent macroscopic flow resistance.\cite{francois2008, francois2009} Hence, the extensional viscosity $\eta_\text{e}$ only quantifies the elongation contribution to fluid stress when averaged over sufficiently large spatial and temporal scales. Recent work has also demonstrated the strong role of polymer elongation on the breakup dynamics of aqueous droplets, indicating that such transient extension likely plays an important role in flows involving fluid interfaces, as in EOR and groundwater remediation.\cite{juarez2011, ingremeau2013, sachdev2016} While these transient effects have typically neglected in models of polymer solution flow through porous media, ongoing work leveraging microfluidics is beginning to shed light on the central influence of unsteady flows in determining macroscopic transport properties. We thus turn our attention to these time-dependent behaviors in the next Section.

\section{Unstable Flow of Polymer Solutions} \label{sec:unsteady}
\subsection{Unstable Flow of Bulk Solutions}\label{sec:unsteady-bulk}
In a heterogeneous porous medium, $\mathrm{Wi}$ is determined by the deformation of the polymers near the curved solid surfaces, and so depends on the local fluid deformation rate. Alternatively, $\mathrm{De}$ depends on the time needed for the polymers to traverse each individual pore. Thus, when both Wi and De are $\gtrsim1$, chain elongation cannot keep up with the changing flow field as a polymer transits through the pore space, leading to the development of an unsteady flow state with continual fluctuations in polymer conformation---further described in \S\ref{sec:unsteady-porous}. 

A similar instability can arise in bulk solution if the polymer chains are forced to flow along highly-curved fluid streamlines; fluctuations in the polymer conformations build up elastic stresses in the solution, leading to sustained spatial and temporal flow fluctuations reminiscent of `conventional' turbulence, which arises at $\textrm{Re}\gg1$. In the inertial case, turbulence arises because the non-linear inertial term in the Navier-Stokes equations destabilizes the steady-state laminar solution. Similarly, in the inertia-free case of polymer solutions at $\textrm{Re}\ll1$, a flow instability analogous to turbulence arises because non-linearities in the constitutive relation between stress and deformation (\S\ref{sec:dimensionless}) destabilize the steady-state laminar solution. Such instabilities were theorized as early as the 1970s\cite{pearson1976, ho1977} and have been catalogued in diverse geometries, including concentric cylinders, rotating disks, and curved channels since the late 1980s.\cite{larson1992, byars1994, casanellas2016, chakraborty2004, galindo2012, galindo2014, haward2016elastic, hulsen2005, larson1992instabilities, muller1989, larson1990, lee1991, li2012, magda1988transition, mckinley1993, mckinley1991, mckinley1995, oliveira2005, oliveira2009, poole2007, pathak2004, ribeiro2014, sen2009, shaqfeh1996, williamson1996, wilson1999, chilcott1988, haward2019} In all cases, an instability is observed for strong enough flow in geometries with a large amount of streamline curvature.

\begin{figure}[h]
	\centering
	\includegraphics[width=6.5 in]{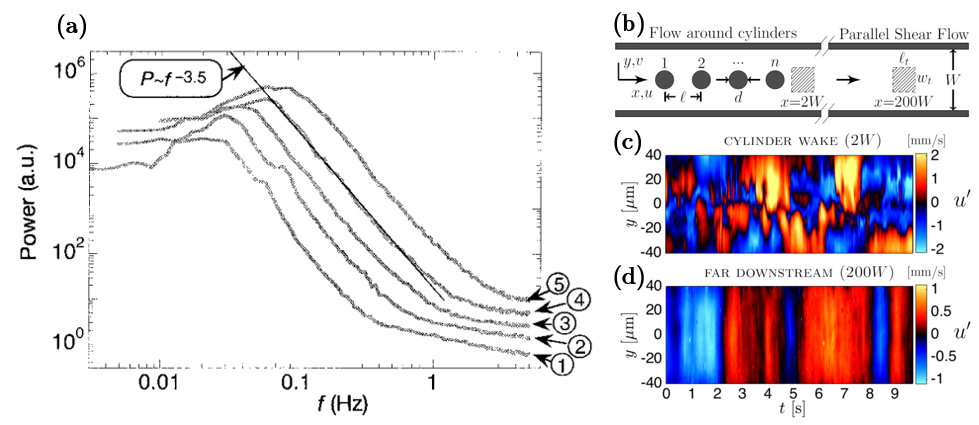}
	\caption{\textbf{a} Power spectra of velocity fluctuations for applied shear rates ($\dot{\gamma}$ = 1.25, 1.85, 2.7, 4, and 5.9 s$^{-1}$ for lines 1--5 respectively) in a cone-plate rheometer.\cite{groisman2000} Copyright 2000, Springer Nature; reprinted with permission. \textbf{b} Schematic of microfluidic porous media, and kymographs of streamwise velocity fluctuations ($u'$) nearby (\textbf{c}) and far downstream (\textbf{d}) of circular pillar array.\cite{qin2017} Copyright 2017, American Physical Society; reprinted with permission.}
	\label{fig:basic_turbulence}
\end{figure}

Groisman and Steinberg gave strong quantitative evidence that this flow instability, even though it arises for $\mathrm{Re}\ll1$, shares many features with high-Re inertial turbulence---hence, this unsteady flow behavior is often termed `elastic turbulence'.\cite{groisman2000} In particular, they showed that the power spectrum of the velocity fluctuations of a polymer solution, sheared at a sufficiently large rate in a cone-plate rheometer, decays as a power law in spatial and temporal frequency (Figure \ref{fig:basic_turbulence} a). Thus, the fluid motion is excited over a broad range of spatial and temporal scales, similar to high-Re turbulence; intriguingly, however, this power law decay (with a -3.5 exponent) has a different value than the -5/3 Kolmogorov exponent characteristic of high-Re turbulence. Simulations are able to recover a similar spectral scaling, supporting the idea that it reflects an elastic instability analogous to turbulence;\cite{berti2008, grilli2013} they also reveal that flow fluctuations are stronger near boundaries,\cite{grilli2013} potentially due to a depletion layer effect.\cite{zami2017} 


Single-molecule imaging of fluorescently-labeled DNA demonstrates that a sharp coil-stretch transition occurs in these unstable flows.\cite{gerashchenko2005} In this case, the mean elongation for polymers in the `stretched' state is found to be approximately 0.85 the contour length.\cite{boffetta2003, liu2014}. Thus, the development of an unstable flow state characterized by sustained flow fluctuations is thought to be linked to the continual elongation of polymer chains. Flow birefringence and single-molecule imaging confirm this expectation. In particular, they demonstrate that flow fluctuations are indeed linked to the continual elongation of polymer chains, which arises for sufficiently large Wi.\cite{haward2013, sousa2015, haward2016elastic, lanzaro2017,jovanovic2011} This picture is further supported by the finding that polymers with more backbone rigidity have suppressed fluctuations, demonstrating that polymer chain stretching is indeed required for this instability.\cite{arratia2006}


Studies of this instability in microfluidic channels shed additional light on its spatial dependence. For example, measurements in a linear channel with straight walls demonstrate that the unstable flow state arises when the fluid is sufficiently perturbed through collisions with an array of pillars. Once sufficiently perturbed, the flow remains unsteady far downstream of the perturbation; this observation suggests the generation of a new dynamical flow state in which fluctuations are sustained indefinitely.\cite{pan2013} However, the statistics of the flow fluctuations are markedly different near the pillars and far downstream.\cite{qin2017}. As the flow progresses downstream, regions of positive or negative streamwise velocity fluctuations coalesce, leading to large rolling advection cell structures (Figure \ref{fig:basic_turbulence} b--d). Moreover, the velocity power spectra show different power-law decays, with a -1.7 exponent near the pillar array and a -2.7 exponent far downstream. These different spectra are concomitant with spatial differences in the measured variation of the shearing and elongational components of the velocity gradient, which mediate polymer stretching. In particular, as the flow progresses downstream, the variation in the elongational component of the velocity gradient dominates, suggesting that the polymer molecules are increasingly stretched by flow gradient in the streamwise direction. These results provide evidence that the unstable state of `elastic turbulence' can develop spatially in parallel channels, suggesting a strong dependence on system geometry. 

In most geometries, the base laminar flow field is characterized by a complex interplay between shear and extension; thus, even predicting the onset of unstable flow can be challenging. Intuitively, one expects that it arises when elastic stresses---characterized by a large value of $\mathrm{Wi}\equiv N_1\left(\dot{\gamma}\right)/2\sigma\left(\dot{\gamma}\right)$---persist over a chain relaxation length scale $\lambda U$ exceeding the effective streamline radius of curvature $\mathscr{R}$. This effective curvature can be thought of as the length scale over which the fluid perturbation persists; thus, the ratio $\lambda U/\mathscr{R}$ is equivalent to the Deborah number De. Thus, one expects that unstable flow arises when the product $\mathrm{Wi}\cdot\mathrm{De}$ is sufficiently large,\cite{pakdel1996} as found in early experimental work.\cite{pearson1976, larson1992} A linear stability analysis of the Upper Convected Maxwell model quantifies this expectation, showing that the largest destabilizing term in the Navier-Stokes equations is of order $$\mathrm{M}\equiv\bigg(\frac{N_1(\dot{\gamma})}{\sigma(\dot{\gamma})}\frac{ \lambda U}{\mathscr{R}}\bigg)^{1/2}$$
\noindent
where in practice, the effective radius $\mathscr{R}$ corresponds to the smallest streamline radius of curvature that elongates polymer chains. This quantity can thus be an order of magnitude smaller than the other characteristic length scales in a given flow geometry,\cite{qin2018} but nevertheless can be empirically calculated from the structural parameters of the system.\cite{haward2016elastic} Unstable flow is thus predicted to arise above a threshold value of this universal parameter, typically called the $\text{M}$ parameter, as defined using the expression above by Pakdel and Mckinley in 1996.\cite{pakdel1996} This expectation was confirmed experimentally for diverse flow geometries;\cite{mckinley1996,haward2016elastic,zilz2012} in practice, the threshold M ranges between $\approx6$ and $12$ in different geometries, and determining exactly why this threshold value varies between experiments is still a topic of current research.  Importantly, polymer solutions applied in EOR and groundwater remediation can have M ranging from $\sim10^{-6}$ to $10^{6}$, indicating that unstable flow can arise in many of these cases.

\subsection{Unstable Flow in Microfluidic Porous Media}\label{sec:unsteady-porous}
Does unstable flow persist in a porous medium? And if so, how does the interplay between unstable flow and pore space geometry impact the flow field and macroscopic flow resistance? Microfluidic investigations over the past two decades have played---and continue to play---a critical role in addressing these questions. Both experiments and complementary simulations have focused on pore space geometries that vary in complexity, ranging from a single contraction, to a single flow obstacle, to a linear channel of obstacles, to a 2D lattice of obstacles, all of which enrich our understanding of unstable flow in porous media. Indeed, unstable flow is expected to arise when high-molecular weight polymers are injected at large speeds; under these conditions, the polymers cannot equilibrate to the continually-changing stresses as they are advected through the tortuous pore space.

\begin{figure}[h]
	\centering
	\includegraphics[width= \textwidth]{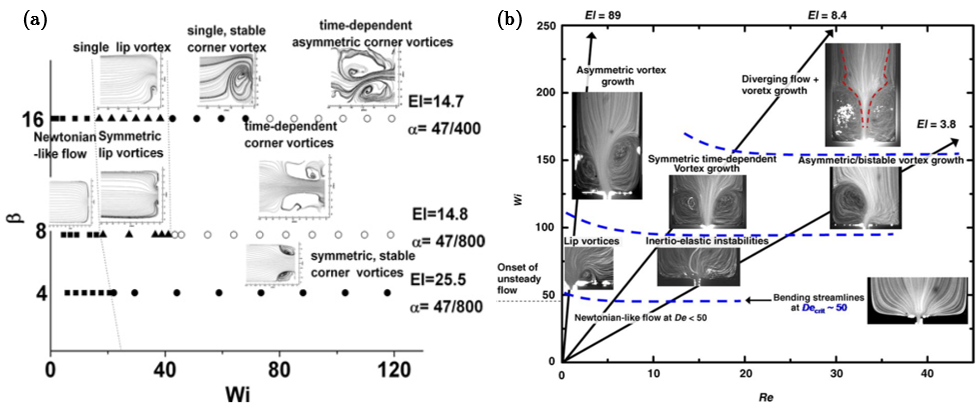}
	\caption{State diagrams indicating the structure of unstable eddies upstream of a sudden constriction at different elasticity numbers, $\text{El}\equiv\text{Wi}/\text{Re}$, and at different constriction ratios $\alpha$, defined as the ratio between the constriction width and the expansion width. \textbf{a} Role of constriction ratio $\beta$ and $\mathrm{Wi}$.\cite{lanzaro2011} Copyright 2011, Elsevier; reprinted with permission. \textbf{b} Role of Re and Wi.\cite{rodd2005} Copyright 2005, Elsevier; reprinted with permission.}
	\label{fig:constriction-ET}
\end{figure}
The simplest model of a pore is a single abrupt constriction. As it enters the constriction, a flowing polymer solution forms large upstream eddies above a threshold value of $\mathrm{M}$. While reminiscent of the steady eddies formed by rigid and elongated polymers (\S\ref{sec:steady-extension} and \S\ref{sec:unsteady-extension}), these eddies fluctuate strongly in time.\cite{koelling1991} The eddy size and asymmetry increases non-linearly with $\mathrm{Wi}$ and with increasing contraction ratio $\beta$, as summarized in the state diagram shown in Figure \ref{fig:constriction-ET}a.\cite{lanzaro2011, lanzaro2011thesis, lanzaro2014, lanzaro2015, lanzaro2017} In one study,\cite{gulati2015} the eddy length $L_\text{e}$ was measured to grow as $L_\text{e}/W\sim\mathrm{Wi}^{0.45}$, where $W$ is the channel width upstream of the constriction, but a universal scaling has not been established. However, similar flow structures can be replicated in simulations,\cite{szabo1997} suggesting that eddy formation is more general. Interestingly, the presence of inertia suppresses temporal fluctuations of these eddies, and promotes symmetry in their structure (Figure \ref{fig:constriction-ET}b).\cite{rodd2005, rodd2007} 

\begin{figure}[h]
	\centering
	\includegraphics[width= \textwidth]{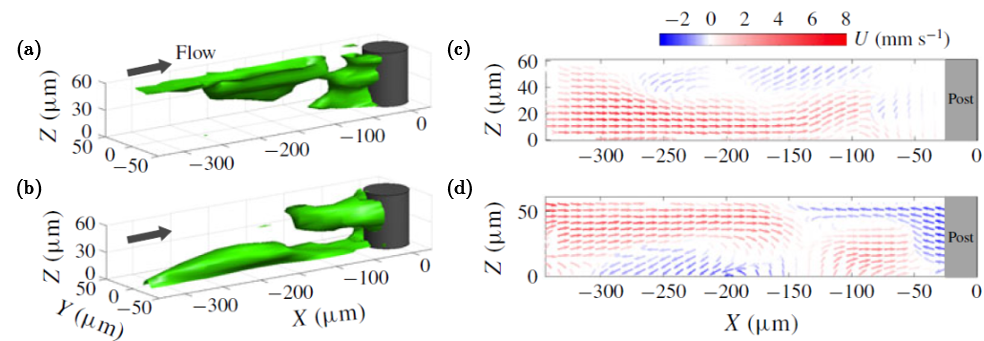}
	\caption{Unstable eddy upstream of a single channel-centered cylinder. The vertical location of the eddy switches between the top and bottom walls over time.\cite{qin2019} Copyright 2019, Cambridge University Press; reprinted with permission.}
	\label{fig:single-pillar}
\end{figure}

\begin{figure}[h]
	\centering
	\includegraphics[width= \textwidth]{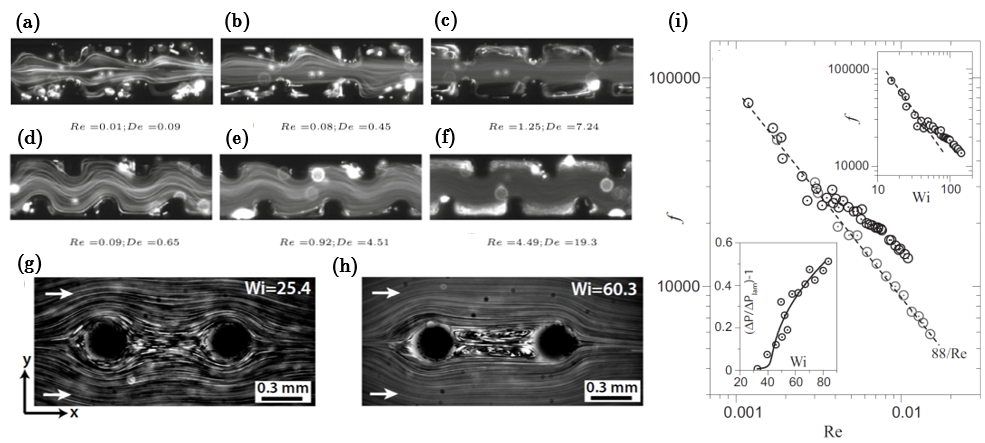}
	\caption{\textbf{(a-h)} Flow structures of unstable polymer solution flow in a channel of square contractions, offset square contractions,\cite{galindo2012} and channel-centered cylinders.\cite{varshney2017} Copyright 2012, Springer; reprinted with permission. \textbf{(i)} Plot of the flow friction factor $f\equiv2D_\text{h}\Delta P/\rho U^2L$, which quantifies flow resistance, for the overall microfluidic channel in the experiment shown in \textbf{(g-h)}; $D_\text{h}$ is the hydraulic radius of the channel,  $\Delta P$ is the pressure drop measured over the total length $L$, $\rho$ is the fluid density, and $U$ is the average channel flow speed. Light grey circles are for a Newtonian fluid while solid black circles are for a solution of 18 MDa polyacrylamide at 100 ppm, corresponding to half the overlap concentration $C^*$. The friction factor departs from the expected Newtonian trend (dashed line) despite the low Re. Insets show that this sudden increase in the flow resistance occurs above a critical Wi corresponding to the onset of the flow instability observed in panels \textbf{g} and \textbf{h}.\cite{varshney2017} Copyright 2017, American Physical Society; reprinted with permission.}
	\label{fig:turbulent-structure}
\end{figure}

Flow through a pore is often represented instead by flow impinging on a channel-centered cylinder. Single-molecule imaging of fluorescently-labeled DNA provides evidence that polymers are stretched and have hysteretic conformations as they flow around the cylinder\cite{francois2009} in a manner similar to coil-stretch hysteresis (\S\ref{sec:unsteady-extension}). Thus, similar to the case of flow through a constriction, upstream eddies form during polymer solution flow around a cylinder in a narrow channel, growing in size and becoming unstable as Wi increases.\cite{qin2019} Holographic imaging enables full characterization of the complex 3D structure of these eddies (Figure \ref{fig:single-pillar}). Intriguingly, the vertical location of the eddy can exhibit discrete switching between the top and bottom walls of the channel,\cite{qin2019} suggesting a possible connection to bistable polymer conformations (described in the context of the coil-stretch transition in \S\ref{sec:unsteady-extension}). Exploring this connection will be a valuable direction for future work. Eddy formation is not general to all such flows: for example, eddies are not observed for polymer solution flows impinging on a cylinder in a wide channel. Instead, an extended wake forms downstream in which the extended chains gradually relax to their equilibrium conformations.\cite{chilcott1988,haward2019,haward2018}

Real-world porous media are typically composed of many interconnected pore expansions and contractions. To more closely mimic this geometry, experiments have investigated polymer solutions flowing through channels with multiple constrictions, channel-centered pillars, and undulating walls arranged in 1D or in 2D arrays. Results obtained for ordered 1D arrays of pores in channels of narrow widths consistently demonstrate the formation of unstable eddies upstream of constrictions, similar to the case of a single constriction (Figure \ref{fig:turbulent-structure}a-h);\cite{galindo2012,khomami1997, arora2002} these eddies also form in flows with moderate inertia ($\mathrm{Re}\sim20$).\cite{kenney2013, shi2015, shi2016} Unstable eddies similarly form in the simple case of a channel with two channel-centered cylinders.\cite{varshney2017} However, when the channel width is made larger, eddies are not observed; instead, an extended wake forms downstream of each cylinder, similar to the case of a single cylinder in a wide channel.\cite{haward2018}

Because polymer stretching is hysteretic and can persist over large length scales,\cite{qin2017} polymer elongation may be retained across multiple pores. This `memory' may therefore provide new spatial structure to the flow over macroscopic scales in a porous medium. Studies of ordered 1D arrays are beginning to reveal such effects. For example, decreasing the distance between pillars in a narrow channel produces stronger fluctuations at similar values of Wi;\cite{shi2016} conversely, in a wider channel, the wake formed downstream of a first cylinder can merge with the wake formed downstream of a second.\cite{haward2018}

This complex coupling between pores also manifests in 2D arrays, showing flow behavior that can differ strongly from the 1D case. For example, polymer solutions flowing through ordered 2D arrays show strong velocity fluctuations throughout the pore space---but eddies are never observed.\cite{howe2015, clarke2016} Intriguingly, in other studies of flow through ordered 2D arrays of circular, square, and triangular pillars, large triangular `dead zones' form on the upstream faces of the pillars; these dead zones are characterized by strong polymer compression and elongation, but unlike eddies, they show no apparent recirculation.\cite{kawale2017b, kawale2019} These dead zones periodically grow and disappear; moreover, the frequency of this process increases further downstream along the porous medium, indicating that coupling between pores influences the flow.\cite{kawale2017a} Consistent with these observations, the temporal variation in pressure measurements increases downstream along ordered 2D arrays of circular pillars, suggesting that instabilities grow spatially.\cite{talwar1995, khomami1997} 

Combining such flow visualization to measurements of the overall pressure drop provides a way to directly link flow structure to macroscopic flow resistance. Such measurements show a dramatic increase in macroscopic flow resistance---mimicking the increase in $\eta_\text{app}$ observed in bulk porous media---at the onset of unstable flow (Figure \ref{fig:turbulent-structure}i).\cite{clarke2016, de2016, varshney2017, de2018viscoelastic} Simulations in model porous geometries can also capture the onset of flow fluctuations\cite{sadanandan2004, de2016coupled, de2017b} and the associated increase in flow resistance.\cite{vazquez2012, de2017a, de2017viscoelastic} These recent findings thus suggest the possibility that the striking increase in polymer solution flow resistance measured in bulk porous media reflects the onset of unstable flow.

\begin{figure}[h]
	\centering
	\includegraphics[width= \textwidth]{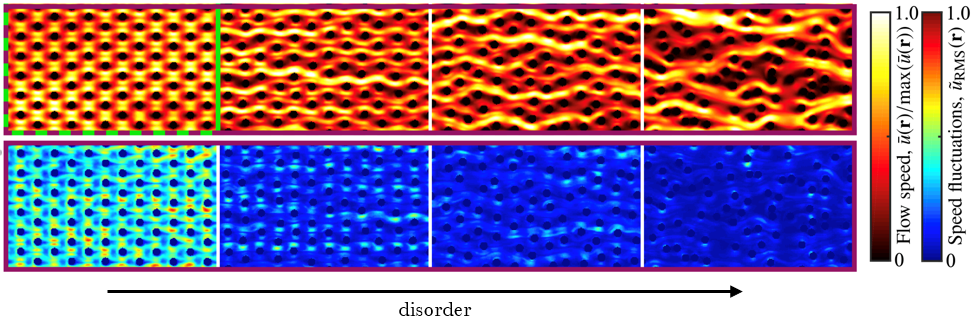}
	\caption{Top row shows normalized time-averaged speed field for $\text{Wi}\approx4$. Bottom row shows corresponding normalized map of speed fluctuations. Left-most column is a perfectly-ordered array, characterized by strong fluctuations; disorder increases moving to the right, showing that disorder promotes the formation of preferential flow channels and suppresses flow fluctuations.\cite{walkama2019} }
	\label{fig:guasto}
\end{figure}

Real-world porous media, however, are typically disordered: the pore sizes and spacing between grains are not uniform. While the majority of experiments and simulations have focused on studying polymer solution flow in ordered arrays, ongoing work is beginning to reveal that structural disorder can dramatically impact flow behavior. For example, flow visualization in 2D arrays of circular pillars shows that while a highly unstable flow state arises in an ordered medium due to sustained polymer elongation, velocity fluctuations are nearly completely suppressed in a disordered array.\cite{walkama2019} Analysis of the pore-scale velocity field suggests an explanation: in a disordered array, interactions between the pores causes the flow to concentrate in finger-like channels dominated by shear, not elongation (Figure \ref{fig:guasto}). This suppression of elongation can be interpreted as a local reduction in $\mathrm{Wi}\equiv N_1\left(\dot{\gamma}\right)/2\sigma\left(\dot{\gamma}\right)$, and thus a local reduction in M, which quantifies flow destabilization. Other 2D experiments indicate a key influence of polymer solution rheology on the formation of these finger-like channels, suggesting that they develop when the fluid is shear-thinning and thus can flow more easily in channels of high shear. Hence, while microfluidic experiments in ordered arrays suggest a tantalizing link between the onset of unstable flow and the increased flow resistance measured in bulk porous media, the influence of pore-scale disorder and fluid rheology complicate this interpretation. Developing a unified understanding of unstable polymer solution flow through porous media---which would enable accurate prediction of the pore-scale flow structure and macroscopic transport behavior---therefore remains an important and outstanding goal for future work.

\section{Improved Fluid Recovery From Porous Media}\label{sec:multiphase}
A prominent application of polymer solutions is to recover trapped non-aqueous fluid from porous media in EOR and groundwater remediation. In these cases, globules of oil or an organic contaminant are trapped in the pore space by capillary forces. A globule of length $L_\text{glob}$ along the flow direction can only be displaced out of the pores when the viscous pressure drop across it, $|\Delta P|=\eta_{\text{app}}(Q/A)L_\text{glob}/k$, exceeds the capillary pressure threshold $\sim\gamma/a$ for squeezing the fluid-fluid interface through the pores, where $\gamma$ is the interfacial tension and $a$ is the pore size.\cite{datta2014b} This criterion is often expressed in terms of the Capillary number, $\mathrm{Ca}\equiv\eta_{\text{app}}(Q/A)/{\gamma}$; displacing trapped non-aqueous fluid globules usually requires $\mathrm{Ca}\geq10^{-4}$, which is higher than is typically accessible in field applications (Table \ref{tab:table1}).\cite{datta2014b, morrow1988, amili2006}

Conventional belief has held that polymer solutions are not able to increase $\mathrm{Ca}$ far above this threshold under typical operating conditions. As a result, dilute solutions of low molecular weight polymers have traditionally been used simply to suppress fingering instabilities during injection and thereby maintain flow uniformity during fluid recovery.\cite{sorbie2013,du2004} However, this view has recently been challenged: EOR field tests with more concentrated solutions of high molecular weight polymers often yield unexpectedly high recoveries of oil fluid far beyond what is expected from a simple estimate of $\mathrm{Ca}$ with $\eta_{\text{app}}$ given by the shear viscosity.\cite{wang2010,wang2011,sandiford1964, pitts1995, wei2014, vermolen2014} These findings have prompted the use of concentrated polymer solutions for groundwater remediation as well.\cite{roote1998, giese2002, darwish2003, smith2008} 

Unfortunately, inconsistencies in the literature make it difficult to define the conditions for improved fluid displacement: many studies report improved oil recovery from sand packs and cores when polymer solutions are injected,\cite{sandiford1964, vermolen2014, durst1981, huh2008} while others report no change or decreased recovery compared to polymer-free water.\cite{schneider1982, giese2002, sandengen2017} Reconciling these inconsistencies is challenging, due to the inability of typical experiments to probe the flow behavior inside the pore space with sufficient spatial and temporal resolution. As a result, the mechanisms by which polymer solutions can enhance fluid recovery remain poorly understood and hotly debated.\cite{zhu2012, urbissinova2010, zhang2011} Hence, microfluidic platforms are increasingly being used to shed light on fluid displacement from porous media.\cite{xu2017,nilsson2013, lacey2017, wang2011}

Several proposed mechanisms rely on a potential influence of polymers on the immiscible fluid interface. For example, it has been speculated that polymer solution viscoelasticity alters the shape of the fluid displacement front in a porous medium.\cite{avendano2013} However, how this process impacts macroscopic fluid recovery is unclear. In another mechanism, it is speculated that the extensional viscosity of polymer solutions suppresses non-aqueous globules from breaking up as they transit through the tortuous pore space. This breakup process---analogous to the Rayleigh-Plateau instability---is thought to arise in strongly disordered pore geometries\cite{roof1970} and results in globules with smaller $L_\text{glob}$. The viscous pressure drop across each globule, and thus its ability to be displaced and recovered, is thus reduced. Due to their extensional viscosity, polymers are thought to suppress this breakup process---as shown in idealized experiments and simulations\cite{derzsi2013, gupta2016, nooranidoost2016, nekouei2017}---and thereby improve non-aqueous fluid recovery. Experiments on bulk rock cores show indirect agreement.\cite{huh2008} However, direct confirmation of this effect inside a porous medium, and quantification of the potential increase in fluid recovery, is lacking. Addressing these questions will be valuable direction for future work.  

Other proposed mechanisms rely on an increase in $\eta_\text{app}$, which therefore increases $\mathrm{Ca}$, potentially above the threshold needed to displace trapped non-aqueous fluid globules. This view is supported by measurements on bulk porous media that directly link an increase in macroscopic flow resistance to enhanced fluid recovery.\cite{wang2011,sandiford1964, pitts1995, wei2014, vermolen2014,clarke2016} 

As described in \S\ref{sec:steady-extension}, one possibility is that polymer adsorption onto the solid matrix partially or fully clogs pores and increases $\eta_\text{app}$.\cite{zaitoun1988, zaitoun1998, darwish2003} While experiments suggest that this mechanism is frequently present, the resultant increase in local flow resistance is often much smaller than the measured macroscopic increase in $\eta_\text{app}$, and thus cannot fully explain enhanced fluid recovery.\cite{grattoni2004} 

As described in \S\ref{sec:viscosity}, another possibility is that the measured increase in $\eta_\text{app}$ reflects an increase in the polymer extensional viscosity. Microfluidic platforms provide a way to directly quantify this effect, revealing that for sufficient extension rates, the extensional viscosity can indeed abruptly increase\cite{kim2018}---potentially providing a sufficient increase in $\eta_\text{app}$ to displace trapped fluids.\cite{haward2003, odell2006} However, quantitative tests of this idea are limited, and future work is needed to determine if the expected increase of $\eta_{app}$ results in a sufficient increase in $\text{Ca}$ to enhance fluid recovery.

\begin{figure}[h]
	\centering
	\includegraphics[width=0.95\textwidth]{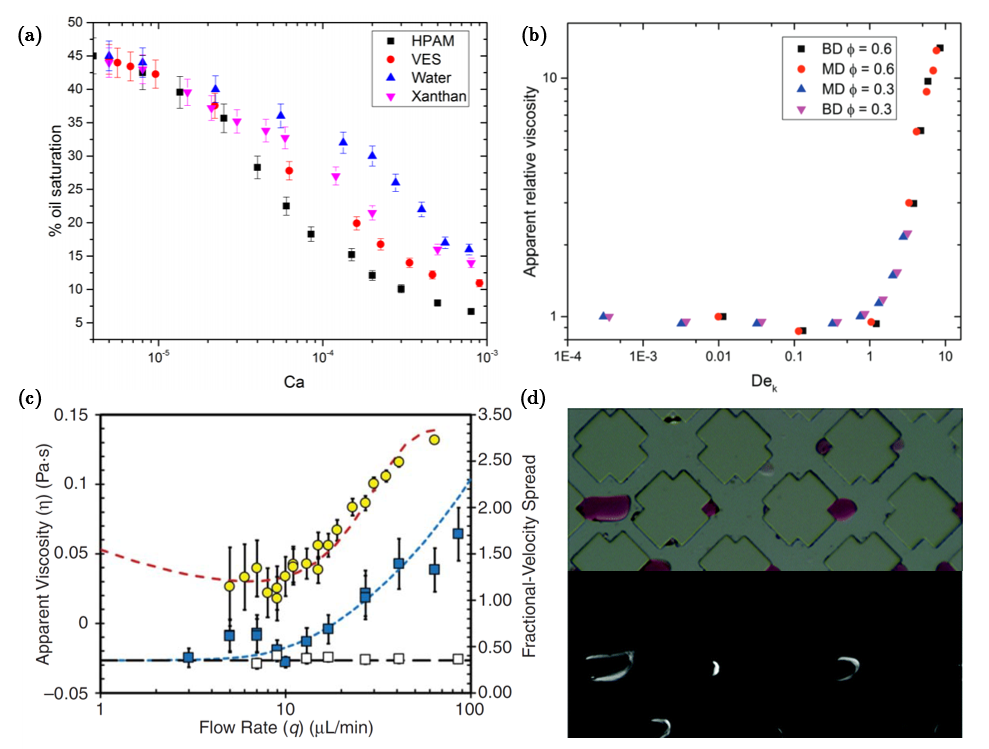}
	\caption{\textbf{a} Reduction in oil saturation at similar apparent capillary numbers Ca by introducing polymers.\cite{de2018viscoelastic} Copyright 2018, Elsevier; reprinted with permission. \textbf{b} Simulations qualitatively match the experimental increase in flow resistance ($\eta_{app}$) at the onset of unstable velocity fluctuations.\cite{de2017viscoelastic} Copyright 2017, Royal Society of Chemistry; reprinted with permission. \textbf{c} Apparent thickening of $\eta_{app}$ (circles) coincides with increased velocity spread for polymer solution flow (filled squares) over Newtonian baseline (open squares).\cite{clarke2015} \textbf{d} The onset of velocity fluctuations and the increase in flow resistence is also coincident with observed fluctuations in the aqueous-nonaqueous interfaces, and an improved displacement of oil globules. Top shows image of oil trapped in microfluidic porous network; bottom shows integrated difference images, which indicate interface fluctuations.\cite{clarke2015} Copyright 2015, Royal Society of Chemistry; reprinted with permission. }
	\label{fig:napl}
\end{figure}

Recent work suggests a key role played by flow instabilities in enhancing fluid recovery. Indeed, microfluidic experiments have shown that unstable flow is often characterized by a dramatic increase in $\eta_\text{app}$, as described in \S\ref{sec:unsteady-porous}---leading to a concomitant increase in $\text{Ca}$.\cite{varshney2017,clarke2016, de2018flow,  de2018viscoelastic,de2017viscoelastic}  This idea has been directly verified using oil globules trapped in a microfluidic porous medium: injecting polymer solutions can increase the apparent $\text{Ca}$ by nearly an order of magnitude, thereby increasing oil recovery (Figure \ref{fig:napl}a-b).\cite{nilsson2013} Independent measurements on different microfluidic porous media show similar results, and demonstrate that the increase in $\text{Ca}$ coincides with an abrupt increase in flow fluctuations, reflecting the onset of unstable flow.\cite{clarke2016} In addition to increasing $\eta_\text{app}$, imaging suggests that these flow fluctuations can also induce fluctuations at the aqueous/non-aqueous fluid contact line with the solid matrix (Figure \ref{fig:napl}d). These fluctuations are speculated to reduce contact line pinning, thereby further promoting trapped globule displacement.\cite{clarke2016, mitchell2016} Together, these results provide compelling evidence that flow instabilities can in some cases contribute to the increase in macroscopic flow resistance, and the concomitant increase in non-aqueous fluid recovery, that has been observed in bulk porous media. Careful assessment of the generality of this phenomenon, and developing the ability to predict the magnitude of increased flow resistance and fluid recovery, will therefore be an important direction for future research.

\section{Outlook}\label{sec:outlook}
We have described how, through precise control over channel geometry, flow conditions, and simultaneous flow visualization, microfluidic technologies shed light on the unusual flow properties of polymer solutions in porous media. Specifically, microfluidic experiments have deepened our understanding of how single-molecule elongation increases steady-state flow resistance and generates an unstable flow state. Together, these effects help to explain macroscopic measurements of increased flow resistance and fluid recovery from bulk porous media. Thus, by providing a fundamental understanding of the underlying physics, these studies help to provide quantitative principles by which polymer solution flows can be predicted and controlled for applications like enhanced oil recovery and groundwater remediation. However, several key questions remain unanswered, which we highlight below as possible directions for future research in this area. 

\textit{Extensional viscosity measurements.} Microfluidic devices with precisely-controlled boundaries now enable the investigation of flows with pure extension, as described in \S\ref{sec:viscosity}. Such platforms provide an opportunity to systematically measure extensional viscosity over different extension rates. While shear viscosity measurements abound, many applications involve the use of polymer solutions under extension. Obtaining measurements of extensional viscosity for different classes of polymers, in different solvents and environmental conditions, will provide key data to guide these applications. Moreover, these data will enable quantitative assessment of the contribution made by extensional viscosity to macroscopic flow resistance measured for bulk porous media. 

\textit{Contribution of eddies to flow resistance.} As described in \S\ref{sec:steady-extension}, eddies often form when polymer solutions are forced through constrictions. These regions of intense recirculation pose an additional source of dissipation and thus contribute to the macroscopic pressure drop. However, a way to quantitatively predict the contribution of eddies to macroscopic flow resistance is lacking. Systematic tests of eddies formed using polymers of different flexibilities and molecular weights could provide crucial data that could help address this gap in knowledge. 

\textit{Transient molecular conformations.} While imaging of single molecules in microfluidic channels (\S\ref{sec:unsteady-extension}) reveals that hysteretic and transient conformations (e.g. dumbbell, kinked, half-dumbbell, folded, end-coiled) impact chain relaxation dynamics, it is unclear whether such transient states arise in flow through more complex porous media, and whether they influence flow behavior. Single-molecule imaging of large polymers in model porous media could provide an answer. 

\textit{Confinement effects.} In highly-confined pore spaces, polymer repulsion from the walls can create a depletion layer.\cite{barnes1995, sochi2011}. This depletion layer arises from a combination of shear-induced migration,\cite{fang2005, ma2005, kekre2010} entropic or steric effects,\cite{joanny1979} and electrostatic interactions.\cite{uematsu2015} Though typically neglected from macroscopic models of flow, the formation of a depletion layer could play a considerable role on transport---for example, possibly suppressing polymer adsorption and leading to viscosity reduction at the pore surfaces.\cite{zami2017} Nanofluidic platforms, which provide a means to test pores of width approaching molecular dimensions, could shed light on these effects. 

\textit{Universal description of unstable flow.} While the onset of unstable flow can be predicted for polymer solutions in bulk media (\S\ref{sec:unsteady-bulk}), how to generally predict this onset in heterogeneous porous media is unknown. The M parameter that universally predicts instability in bulk flows incorporates the effective streamline curvature $\mathscr{R}$; however, it is unclear how to calculate $\mathscr{R}$ to accurately describe a highly tortuous pore space. 

Most experiments are reported in terms of the Weissenberg number Wi, and often use the approximate form $\mathrm{Wi}\equiv\lambda \dot{\gamma}$, with differing choices of $\lambda$ and how to calculate $\dot{\gamma}$. As a result, a wide range of different flow behaviors is reported. Moreover, while recent experiments suggest that unstable flow can develop spatially due to coupling between flow in different pores, an understanding of this behavior is lacking. An open challenge is to develop a unified description of these diverse flow behaviors, capable of predicting the unstable flow that arises for different polymer solutions, pore space geometries, and imposed flow conditions. 

\textit{Different fluid displacement mechanisms.} As described in \S\ref{sec:multiphase}, numerous mechanisms have been identified to explain how polymer solutions can enhance recovery of trapped non-aqueous fluid from porous media. However, quantitative and predictive models of fluid recovery remain lacking. Studies typically focus on describing one mechanism, while in reality multiple mechanisms may be at play---thus hindering a unified understanding of fluid recovery. Thus, an important direction for future work will be to systematically assess the occurrence and relative influence of the different mechanisms identified thus far (suppressed fluid break up, permeability reduction through polymer adsorption, the contribution of extensional viscosity, enhanced dissipation and contact line fluctuations due to unstable flow) under different flow conditions.

\textit{Pore-scale disorder.} While the majority of experiments and simulations have focused on studying polymer solution flow in ordered arrays, real-world porous media are typically disordered. Recent work has shown that disorder may strongly alter flow behavior---in some cases, completely suppressing the onset of unstable flow. However, predicting when this suppression arises---i.e. how sensitive it is to the pore-space geometry---and why it arises is still an open challenge. 

\textit{Microfluidic surface chemistry.} Most microfluidic visualization studies use devices made with PDMS, which may not adequately recapitulate the complex surface chemistry of real-world porous media. However, advances continue to be made in the development of microfluidic channels etched directly into mineral substrates, thus enabling flow visualization in channels with surfaces made of real rock material.\cite{zhang2010, song2014, singh2017, gerold2018} While this technology has been used to shed light on questions relating to chemical reactions in porous media, it has yet to be employed in studies of polymer solution flow.

\textit{Fluid rheology.} While some simulations suggest that shear-thinning effects do not influence polymer solution flow, other work suggests a dominant role for the fluid shear rheology. For example, shear-thinning is speculated to promote the formation of preferential flow channels, which reduce polymer elongation. Thus, developing a quantitative description of how fluid rheology impacts the flow behaviors described in this paper is an important direction for future work.

\textit{Pore space dimensionality.} The pore space of a 2D medium is considerably less connected and tortuous than that of a disordered 3D medium. This difference can lead to dramatic differences in polymer solution flow behavior and fluid recovery; thus, the applicability of experiments performed on 2D microfluidic devices to 3D porous media is unclear. Do the flow behaviors seen in 2D arise in 3D media as well? Emerging microfluidic techniques enable us to now answer this question---for example, using novel 3D fabrication methods\cite{xu2017} or using refractive index-matching of model 3D media.\cite{datta2013a,krummel2013,datta2014a,datta2014b}

\textit{Mixing and solute transport.} The velocity fluctuations that characterize unstable flow of polymer solutions can also enhance mixing and solute dispersion on mesoscopic scales.\cite{larson1992,groisman2001, groisman2004,qin2018,scholz2014,babayekhorasani2016, aramideh2019} This enhanced mixing could improve EOR and groundwater remediation efforts by improving the transport of key additives like surfactants, colloids, and oxidants. However, systematic tests of this behavior are lacking. 

Enhanced mixing can be leveraged in other industrial processes as well. Mixing in microfluidic channels is typically diffusion-limited because of the low Reynolds numbers, but elastic flow instabilities have been shown to dramatically improve the speed of mixing.\cite{burghelea2004, lam2009} Similarly, this enhanced mixing can be leveraged for improved heat transfer in other industrial processes that require low Reynolds numbers.\cite{ligrani2017}

\textit{Other applications.} Most studies focus on polymer solution flows in the context of oil recovery and groundwater remediation. However, the work described in this review article may also be relevant to other applications that seek to use polymer solutions in porous media: filtration,\cite{bourgeat2003} fabrication of new materials, thin film processing,\cite{diao2014} extrusion of polymers during 3D-printing,\cite{compton2014, wang20173d} and novel forms of chromatography.\cite{luo1996} Extending previous results to these new settings will be a useful direction for future research. 

While we have focused on flows with $\mathrm{Re}\ll1$, many of these applications may involve $\text{Re}\geq1$. How do the results obtained thus far extend to this new flow regime? Previous experiments shed some light on this question. Studies of flow through 1D arrays of posts shows that unstable upstream eddies form, similar to the low-Re case, when $\mathrm{Re}\sim20$.\cite{kenney2013, shi2015, shi2016} However, other experiments suggest that inertia suppresses temporal fluctuations of these eddies, and promotes symmetry in their structure; more generally, inertia appears to suppress the onset of elastic instabilities,\cite{rodd2005,rodd2007, lanzaro2011thesis} in part due to scission of polymers.\cite{vanapalli2006}

\section{Acknowledgements}
It is a pleasure to acknowledge P. D. Olmsted, R. K. Prud'homme, and H. A. Stone for stimulating discussions. Acknowledgment is made to the Donors of the American Chemical Society Petroleum Research Fund for partial support of this research. This material is also based upon work supported by the National Science Foundation Graduate Research Fellowship Program (to C.A.B.) under Grant No. DGE-1656466. Any opinions, findings, and conclusions or recommendations expressed in this material are those of the authors and do not necessarily reflect the views of the National Science Foundation. C.A.B. was supported in part by the Mary and Randall Hack Graduate Award of the Princeton Environmental Institute. A.S. was supported in part by the Lidow Thesis Fund at Princeton University and the Dede T. Bartlett P03 Fund for Student Research through the Andlinger Center for Energy and the Environment.

\newpage

\begin{figure}[h]
	\centering
	\includegraphics[width=3.125in]{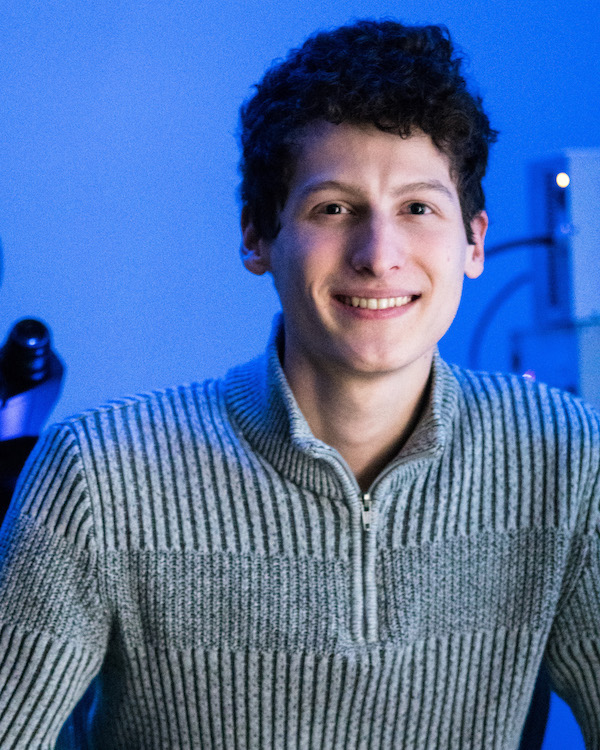}
	\caption*{\textbf{Christopher Browne} is a Ph.D. candidate and NSF Graduate Research Fellow in Chemical and Biological Engineering at Princeton University. His research investigates how the unstable flow of polymer solutions develops in complex three-dimensional porous media. He earned his B.S. in Chemical Engineering from Purdue University.}
	\label{fig:chris}
\end{figure}
\newpage

\begin{figure}[h]
	\centering
	\includegraphics[width=3.125in]{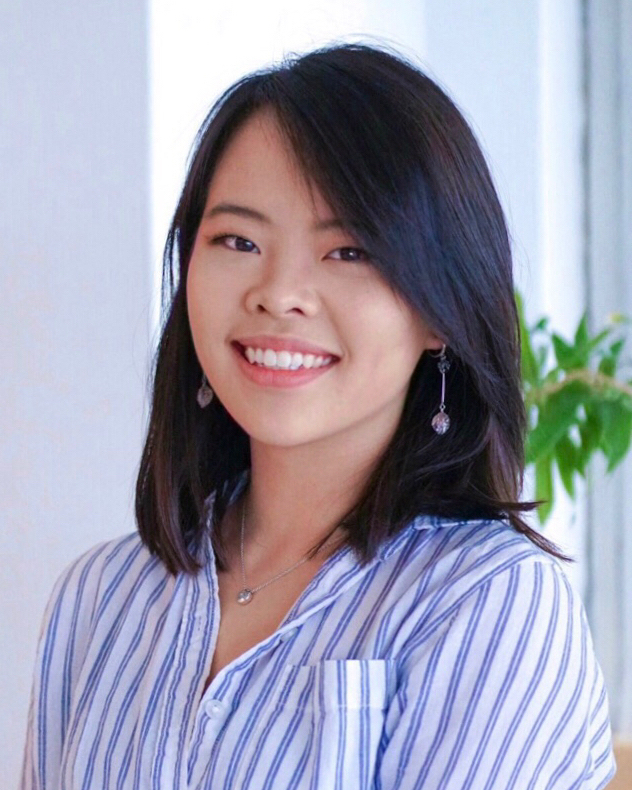}
	\caption*{\textbf{Audrey Shih} is a senior undergraduate at Princeton University pursuing a Bachelor's degree in Chemical and Biological Engineering and a certificate in Materials Science and Engineering. Her current research focuses on the influence of porous media geometry on polymer flow by using microfluidics to examine the link between chain conformations and macroscopic flow structure.}
	\label{fig:audrey}
\end{figure}
\newpage

\begin{figure}[h]
	\centering
	\includegraphics[width=3.125in]{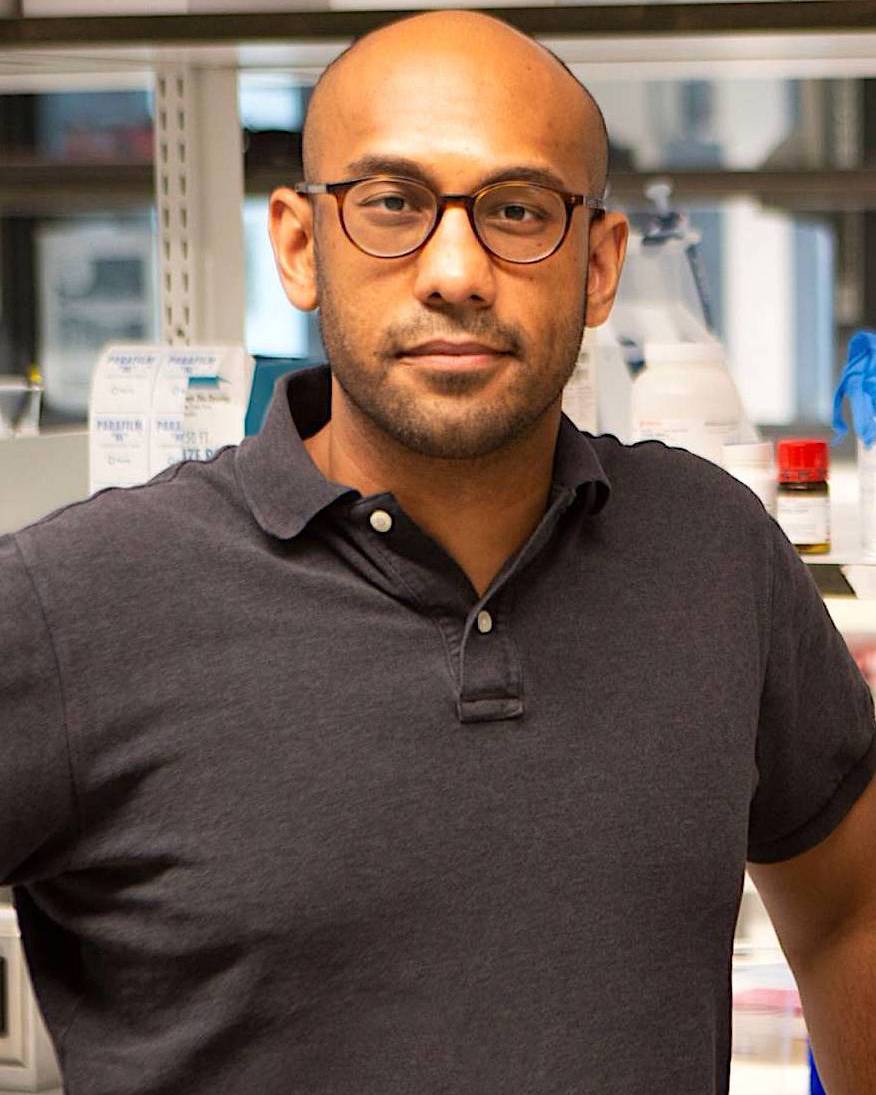}
	\caption*{\textbf{Sujit Datta} is an Assistant Professor of Chemical and Biological Engineering at Princeton University. His lab's research focuses on the behavior of soft and active materials in complex settings, motivated by challenges like developing cleaner oil/gas recovery, more effective water remediation, and targeted drug delivery. Prof. Datta is the recipient of the LeRoy Apker Award from the American Physical Society, the Andreas Acrivos Award in Fluid Dynamics from the American Physical Society, the ACS Petroleum Research Fund New Investigator Award, the Alfred Rheinstein Faculty Award, and multiple Princeton Engineering Commendations for Outstanding Teaching.}
	\label{fig:sujit}
\end{figure}
\newpage

\bibliography{ms}

\begin{thebibliography}{100}

\bibitem{sorbie2013}
Kenneth~S Sorbie.
\newblock {\em Polymer-improved oil recovery}.
\newblock Springer Science \& Business Media, 2013.

\bibitem{roote1998}
DS~Roote.
\newblock Technology status report: in situ flushing.
\newblock {\em Ground Water Remediation Technology Analysis Center (available
  at http://www. gwrtac. org)}, 1998.

\bibitem{liu2007surfactant}
Qiang Liu, Mingzhe Dong, Shanzhou Ma, and Yun Tu.
\newblock Surfactant enhanced alkaline flooding for western canadian heavy oil
  recovery.
\newblock {\em Colloids and Surfaces A: Physicochemical and Engineering
  Aspects}, 293(1-3):63--71, 2007.

\bibitem{yekeen2017}
Nurudeen Yekeen, Ahmad~Kamal Idris, Muhammad~A Manan, Ali~Mohamed Samin,
  Abdul~Rahim Risal, and Tan~Xin Kun.
\newblock Bulk and bubble-scale experimental studies of influence of
  nanoparticles on foam stability.
\newblock {\em Chinese Journal of Chemical Engineering}, 25(3):347--357, 2017.

\bibitem{samanta2012}
Abhijit Samanta, Achinta Bera, Keka Ojha, and Ajay Mandal.
\newblock Comparative studies on enhanced oil recovery by alkali--surfactant
  and polymer flooding.
\newblock {\em Journal of Petroleum Exploration and Production Technology},
  2(2):67--74, 2012.

\bibitem{gbadamosi2019}
Afeez~O Gbadamosi, Radzuan Junin, Muhammad~A Manan, Nurudeen Yekeen, and Agi
  Augustine.
\newblock Hybrid suspension of polymer and nanoparticles for enhanced oil
  recovery.
\newblock {\em Polymer Bulletin}, pages 1--38, 2019.

\bibitem{raffa2016}
Patrizio Raffa, Antonius~A Broekhuis, and Francesco Picchioni.
\newblock Polymeric surfactants for enhanced oil recovery: A review.
\newblock {\em Journal of Petroleum Science and Engineering}, 145:723--733,
  2016.

\bibitem{james1975}
David~F James and DR~McLaren.
\newblock The laminar flow of dilute polymer solutions through porous media.
\newblock {\em Journal of Fluid Mechanics}, 70(4):733--752, 1975.

\bibitem{wang2011}
Demin Wang, Gang Wang, Huifen Xia, et~al.
\newblock Large scale high visco-elastic fluid flooding in the field achieves
  high recoveries.
\newblock In {\em SPE Enhanced Oil Recovery Conference}. Society of Petroleum
  Engineers, 2011.

\bibitem{sandiford1964}
BB~Sandiford et~al.
\newblock Laboratory and field studies of water floods using polymer solutions
  to increase oil recoveries.
\newblock {\em Journal of Petroleum Technology}, 16(08):917--922, 1964.

\bibitem{pitts1995}
Malcolm~J Pitts, Tom~A Campbell, Harry Surkalo, Kon Wyatt, et~al.
\newblock Polymer flood of the rapdan pool, saskatchewan, canada.
\newblock {\em SPE Reservoir Engineering}, 10(03):183--186, 1995.

\bibitem{wei2014}
Bing Wei, Laura Romero-Zer{\'o}n, and Denis Rodrigue.
\newblock Oil displacement mechanisms of viscoelastic polymers in enhanced oil
  recovery (eor): a review.
\newblock {\em Journal of Petroleum Exploration and Production Technology},
  4(2):113--121, 2014.

\bibitem{vermolen2014}
ECM Vermolen, MJT Van~Haasterecht, SK~Masalmeh, et~al.
\newblock A systematic study of the polymer visco-elastic effect on residual
  oil saturation by core flooding.
\newblock In {\em SPE EOR Conference at Oil and Gas West Asia}. Society of
  Petroleum Engineers, 2014.

\bibitem{clarke2016}
Andrew Clarke, Andrew~M Howe, Jonathan Mitchell, John Staniland, Laurence~A
  Hawkes, et~al.
\newblock How viscoelastic-polymer flooding enhances displacement efficiency.
\newblock {\em SPE Journal}, 21(03):675--687, 2016.

\bibitem{du2004}
Y~Du, L~Guan, et~al.
\newblock Field-scale polymer flooding: lessons learnt and experiences gained
  during past 40 years.
\newblock In {\em SPE International Petroleum Conference in Mexico}. Society of
  Petroleum Engineers, 2004.

\bibitem{haward2003}
Simon~J Haward and Jeffrey~A Odell.
\newblock Viscosity enhancement in non-newtonian flow of dilute polymer
  solutions through crystallographic porous media.
\newblock {\em Rheologica acta}, 42(6):516--526, 2003.

\bibitem{odell2006}
JA~Odell and SJ~Haward.
\newblock Viscosity enhancement in non-newtonian flow of dilute aqueous polymer
  solutions through crystallographic and random porous media.
\newblock {\em Rheologica acta}, 45(6):853--863, 2006.

\bibitem{zaitoun1988}
A~Zaitoun, N~Kohler, et~al.
\newblock Two-phase flow through porous media: effect of an adsorbed polymer
  layer.
\newblock In {\em SPE Annual Technical Conference and Exhibition}. Society of
  Petroleum Engineers, 1988.

\bibitem{zaitoun1998}
A~Zaitoun, H~Bertin, D~Lasseux, et~al.
\newblock Two-phase flow property modifications by polymer adsorption.
\newblock In {\em SPE/DOE improved oil recovery symposium}. Society of
  Petroleum Engineers, 1998.

\bibitem{huh2008}
Chun Huh, Gary~Arnold Pope, et~al.
\newblock Residual oil saturation from polymer floods: laboratory measurements
  and theoretical interpretation.
\newblock In {\em SPE Symposium on Improved Oil Recovery}. Society of Petroleum
  Engineers, 2008.

\bibitem{pearson1976}
JRA Pearson.
\newblock Instability in non-newtonian flow.
\newblock {\em Annual Review of Fluid Mechanics}, 8(1):163--181, 1976.

\bibitem{larson1992}
RG~Larson.
\newblock Flow-induced mixing, demixing, and phase transitions in polymeric
  fluids.
\newblock {\em Rheologica Acta}, 31(6):497--520, 1992.

\bibitem{groisman2000}
Alexander Groisman and Victor Steinberg.
\newblock Elastic turbulence in a polymer solution flow.
\newblock {\em Nature}, 405(6782):53, 2000.

\bibitem{tsiklauri1994}
MG~Tsiklauri.
\newblock Conditions for the occurrence of elastic turbulence in polymer
  solution flows through porous media.
\newblock {\em Journal of engineering physics and thermophysics},
  66(3):233--239, 1994.

\bibitem{mitchell2016}
Jonathan Mitchell, Kyle Lyons, Andrew~M Howe, and Andrew Clarke.
\newblock Viscoelastic polymer flows and elastic turbulence in
  three-dimensional porous structures.
\newblock {\em Soft Matter}, 12(2):460--468, 2016.

\bibitem{blunt2017}
Martin~J Blunt.
\newblock {\em Multiphase flow in permeable media: A pore-scale perspective}.
\newblock Cambridge University Press, 2017.

\bibitem{seymour1997}
Joseph~D Seymour and Paul~T Callaghan.
\newblock Generalized approach to nmr analysis of flow and dispersion in porous
  media.
\newblock {\em AIChE Journal}, 43(8):2096--2111, 1997.

\bibitem{li2017}
Ming Li, Laura Romero-Zer{\'o}n, Florin Marica, and Bruce~J Balcom.
\newblock Polymer flooding enhanced oil recovery evaluated with magnetic
  resonance imaging and relaxation time measurements.
\newblock {\em Energy \& Fuels}, 31(5):4904--4914, 2017.

\bibitem{brown2017}
Jennifer~R Brown, Jacob Trudnowski, Elmira Nybo, Katherine~E Kent, Thomas Lund,
  and Amanda Parsons.
\newblock Quantification of non-newtonian fluid dynamics of a wormlike micelle
  solution in porous media with magnetic resonance.
\newblock {\em Chemical Engineering Science}, 173:145--152, 2017.

\bibitem{shattuck1997}
MD~Shattuck, RP~Behringer, GA~Johnson, and John~G Georgiadis.
\newblock Convection and flow in porous media. part 1. visualization by
  magnetic resonance imaging.
\newblock {\em Journal of Fluid Mechanics}, 332:215--245, 1997.

\bibitem{kutsovsky1996}
YE~Kutsovsky, LE~Scriven, HT~Davis, and BE~Hammer.
\newblock Nmr imaging of velocity profiles and velocity distributions in bead
  packs.
\newblock {\em Physics of Fluids}, 8(4):863--871, 1996.

\bibitem{lebon1996}
L~Lebon, L~Oger, J~Leblond, JP~Hulin, NS~Martys, and LM~Schwartz.
\newblock Pulsed gradient nmr measurements and numerical simulation of flow
  velocity distribution in sphere packings.
\newblock {\em Physics of Fluids}, 8(2):293--301, 1996.

\bibitem{lebon1997}
L~Lebon, J~Leblond, and JP~Hulin.
\newblock Experimental measurement of dispersion processes at short times using
  a pulsed field gradient nmr technique.
\newblock {\em Physics of Fluids}, 9(3):481--490, 1997.

\bibitem{sederman1997}
AJ~Sederman, ML~Johns, AS~Bramley, P~Alexander, and LF~Gladden.
\newblock Magnetic resonance imaging of liquid flow and pore structure within
  packed beds.
\newblock {\em Chemical Engineering Science}, 52(14):2239--2250, 1997.

\bibitem{sederman2001}
AJ~Sederman and LF~Gladden.
\newblock Mri as a probe of the deposition of solid fines in a porous medium.
\newblock {\em Magnetic resonance imaging}, 19(3-4):565--567, 2001.

\bibitem{boger1987}
DV~Boger.
\newblock Viscoelastic flows through contractions.
\newblock {\em Annual review of fluid mechanics}, 19(1):157--182, 1987.

\bibitem{lipscomb1988}
GG~Lipscomb~II, Morton~M Denn, DU~Hur, and David~V Boger.
\newblock The flow of fiber suspensions in complex geometries.
\newblock {\em Journal of Non-Newtonian Fluid Mechanics}, 26(3):297--325, 1988.

\bibitem{mongruel1995}
A~Mongruel and M~Cloitre.
\newblock Extensional flow of semidilute suspensions of rod-like particles
  through an orifice.
\newblock {\em Physics Of Fluids}, 7(11):2546--2552, 1995.

\bibitem{mongruel2003}
A~Mongruel and M~Cloitre.
\newblock Axisymmetric orifice flow for measuring the elongational viscosity of
  semi-rigid polymer solutions.
\newblock {\em Journal of non-newtonian fluid mechanics}, 110(1):27--43, 2003.

\bibitem{zhu2012}
Huaijiang Zhu, Jianhui Luo, Oldoerp Klaus, and Yongzhong Fan.
\newblock The impact of extensional viscosity on oil displacement efficiency in
  polymer flooding.
\newblock {\em Colloids and Surfaces A: Physicochemical and Engineering
  Aspects}, 414:498--503, 2012.

\bibitem{urbissinova2010}
Tolkynay Urbissinova, Japan~J Trivedi, Ergun Kuru, et~al.
\newblock Effect of elasticity during viscoelastic polymer flooding-a possible
  mechanism of increasing the sweep efficiency.
\newblock In {\em SPE western regional meeting}. Society of Petroleum
  Engineers, 2010.

\bibitem{zhang2011}
Zhen Zhang, Jiachun Li, and Jifu Zhou.
\newblock Microscopic roles of “viscoelasticity” in hpma polymer flooding
  for eor.
\newblock {\em Transport in porous media}, 86(1):199--214, 2011.

\bibitem{muggeridge2014}
Ann Muggeridge, Andrew Cockin, Kevin Webb, Harry Frampton, Ian Collins, Tim
  Moulds, and Peter Salino.
\newblock Recovery rates, enhanced oil recovery and technological limits.
\newblock {\em Philosophical Transactions of the Royal Society A: Mathematical,
  Physical and Engineering Sciences}, 372(2006):20120320, 2014.

\bibitem{anbari2018}
Alimohammad Anbari, Hung-Ta Chien, Sujit~S Datta, Wen Deng, David~A Weitz, and
  Jing Fan.
\newblock Microfluidic model porous media: fabrication and applications.
\newblock {\em Small}, 14(18):1703575, 2018.

\bibitem{oconnell2019}
Margaret~G O’Connell, Nancy~B Lu, Christopher~A Browne, and Sujit~S Datta.
\newblock Cooperative size sorting of deformable particles in porous media.
\newblock {\em Soft matter}, 15(17):3620--3626, 2019.

\bibitem{thielicke2014}
William Thielicke and Eize Stamhuis.
\newblock Pivlab--towards user-friendly, affordable and accurate digital
  particle image velocimetry in matlab.
\newblock {\em Journal of Open Research Software}, 2(1), 2014.

\bibitem{blunt2013}
Martin~J Blunt, Branko Bijeljic, Hu~Dong, Oussama Gharbi, Stefan Iglauer,
  Peyman Mostaghimi, Adriana Paluszny, and Christopher Pentland.
\newblock Pore-scale imaging and modelling.
\newblock {\em Advances in Water Resources}, 51:197--216, 2013.

\bibitem{rodd2007}
LE~Rodd, JJ~Cooper-White, DV~Boger, and GH~McKinley.
\newblock Role of the elasticity number in the entry flow of dilute polymer
  solutions in micro-fabricated contraction geometries.
\newblock {\em Journal of Non-Newtonian Fluid Mechanics}, 143(2-3):170--191,
  2007.

\bibitem{kenney2013}
Stephen Kenney, Kade Poper, Ganesh Chapagain, and Gordon~F Christopher.
\newblock Large deborah number flows around confined microfluidic cylinders.
\newblock {\em Rheologica Acta}, 52(5):485--497, 2013.

\bibitem{kawale2017b}
Durgesh Kawale, Gelmer Bouwman, Shaurya Sachdev, Pacelli~LJ Zitha, Michiel~T
  Kreutzer, William~R Rossen, and Pouyan~E Boukany.
\newblock Polymer conformation during flow in porous media.
\newblock {\em Soft matter}, 13(46):8745--8755, 2017.

\bibitem{nilsson2013}
Michael~A Nilsson, Ruta Kulkarni, Lauren Gerberich, Ryan Hammond, Rohitashwa
  Singh, Elizabeth Baumhoff, and Jonathan~P Rothstein.
\newblock Effect of fluid rheology on enhanced oil recovery in a microfluidic
  sandstone device.
\newblock {\em Journal of Non-Newtonian Fluid Mechanics}, 202:112--119, 2013.

\bibitem{ober2013}
Thomas~J Ober, Simon~J Haward, Christopher~J Pipe, Johannes Soulages, and
  Gareth~H McKinley.
\newblock Microfluidic extensional rheometry using a hyperbolic contraction
  geometry.
\newblock {\em Rheologica Acta}, 52(6):529--546, 2013.

\bibitem{graessley2008}
William~Walter Graessley.
\newblock {\em Polymeric liquids and networks: dynamics and rheology}.
\newblock Garland Science, 2008.

\bibitem{balhoff2006}
Matthew~T Balhoff and Karsten~E Thompson.
\newblock A macroscopic model for shear-thinning flow in packed beds based on
  network modeling.
\newblock {\em Chemical Engineering Science}, 61(2):698--719, 2006.

\bibitem{zami2016}
Fr{\'e}d{\'e}ric Zami-Pierre, R~De~Loubens, Michel Quintard, and Yohan Davit.
\newblock Transition in the flow of power-law fluids through isotropic porous
  media.
\newblock {\em Physical review letters}, 117(7):074502, 2016.

\bibitem{machado2016}
Ana{\"\i}s Machado, Hugues Bodiguel, Julien Beaumont, G{\'e}rald Clisson, and
  Annie Colin.
\newblock Extra dissipation and flow uniformization due to elastic
  instabilities of shear-thinning polymer solutions in model porous media.
\newblock {\em Biomicrofluidics}, 10(4):043507, 2016.

\bibitem{batchelor1971}
GK~Batchelor.
\newblock The stress generated in a non-dilute suspension of elongated
  particles by pure straining motion.
\newblock {\em Journal of Fluid Mechanics}, 46(4):813--829, 1971.

\bibitem{dobrynin1995}
Andrey~V Dobrynin, Ralph~H Colby, and Michael Rubinstein.
\newblock Scaling theory of polyelectrolyte solutions.
\newblock {\em Macromolecules}, 28(6):1859--1871, 1995.

\bibitem{turkoz2018}
Emre Turkoz, Antonio Perazzo, Craig~B Arnold, and Howard~A Stone.
\newblock Salt type and concentration affect the viscoelasticity of
  polyelectrolyte solutions.
\newblock {\em Applied Physics Letters}, 112(20):203701, 2018.

\bibitem{kawale2017a}
Durgesh Kawale, Esteban Marques, Pacelli~LJ Zitha, Michiel~T Kreutzer,
  William~R Rossen, and Pouyan~E Boukany.
\newblock Elastic instabilities during the flow of hydrolyzed polyacrylamide
  solution in porous media: Effect of pore-shape and salt.
\newblock {\em Soft matter}, 13(4):765--775, 2017.

\bibitem{fuller1987}
Gerald~G Fuller, Cheryl~A Cathey, Brent Hubbard, and Beth~E Zebrowski.
\newblock Extensional viscosity measurements for low-viscosity fluids.
\newblock {\em Journal of Rheology}, 31(3):235--249, 1987.

\bibitem{cathey1988}
Cheryl~A Cathey and Gerald~G Fuller.
\newblock Uniaxial and biaxial extensional viscosity measurements of dilute and
  semi-dilute solutions of rigid rod polymers.
\newblock {\em Journal of non-newtonian fluid mechanics}, 30(2-3):303--316,
  1988.

\bibitem{jimenez2018}
Leidy~Nallely Jimenez, Jelena Dinic, Nikhila Parsi, and Vivek Sharma.
\newblock Extensional relaxation time, pinch-off dynamics, and printability of
  semidilute polyelectrolyte solutions.
\newblock {\em Macromolecules}, 51(14):5191--5208, 2018.

\bibitem{walter2019}
Anna~V Walter, Leidy~N Jimenez, Jelena Dinic, Vivek Sharma, and Kendra~A Erk.
\newblock Effect of salt valency and concentration on shear and extensional
  rheology of aqueous polyelectrolyte solutions for enhanced oil recovery.
\newblock {\em Rheologica Acta}, 58(3-4):145--157, 2019.

\bibitem{del2017}
Francesco Del~Giudice, Simon~J Haward, and Amy~Q Shen.
\newblock Relaxation time of dilute polymer solutions: A microfluidic approach.
\newblock {\em Journal of Rheology}, 61(2):327--337, 2017.

\bibitem{haward2016microfluidic}
SJ~Haward.
\newblock Microfluidic extensional rheometry using stagnation point flow.
\newblock {\em Biomicrofluidics}, 10(4):043401, 2016.

\bibitem{juarez2011}
Gabriel Juarez and Paulo~E Arratia.
\newblock Extensional rheology of dna suspensions in microfluidic devices.
\newblock {\em Soft Matter}, 7(19):9444--9452, 2011.

\bibitem{ingremeau2013}
Fran{\c{c}}ois Ingremeau and Hamid Kellay.
\newblock Stretching polymers in droplet-pinch-off experiments.
\newblock {\em Physical Review X}, 3(4):041002, 2013.

\bibitem{sachdev2016}
Shaurya Sachdev, Aswin Muralidharan, and Pouyan~E Boukany.
\newblock Molecular processes leading to “necking” in extensional flow of
  polymer solutions: Using microfluidics and single dna imaging.
\newblock {\em Macromolecules}, 49(24):9578--9585, 2016.

\bibitem{kim2018}
Seo~Gyun Kim, Chang~Min Ok, and Heon~Sang Lee.
\newblock Steady-state extensional viscosity of a linear polymer solution using
  a differential pressure extensional rheometer on a chip.
\newblock {\em Journal of Rheology}, 62(5):1261--1270, 2018.

\bibitem{bird1987}
R~Byron Bird, Robert~C Armstrong, and Ole Hassager.
\newblock {\em Dynamics of polymeric liquids. Volume 1: fluid mechanics}.
\newblock A Wiley-Interscience Publication, John Wiley \& Sons, 1987.

\bibitem{peterlin1961}
A~Peterlin.
\newblock Streaming birefringence of soft linear macromolecules with finite
  chain length.
\newblock {\em Polymer}, 2:257--264, 1961.

\bibitem{weissenberg1947}
K~Weissenberg.
\newblock A continuum theory of rhelogical phenomena, 1947.

\bibitem{howe2015}
Andrew~M Howe, Andrew Clarke, and Daniel Giernalczyk.
\newblock Flow of concentrated viscoelastic polymer solutions in porous media:
  effect of mw and concentration on elastic turbulence onset in various
  geometries.
\newblock {\em Soft Matter}, 11(32):6419--6431, 2015.

\bibitem{macosko1994}
Christopher~W Macosko and Ronald~G Larson.
\newblock {\em Rheology: principles, measurements, and applications}.
\newblock Vch New York, 1994.

\bibitem{dealy2010}
JM~Dealy.
\newblock Weissenberg and deborah numbers—their definition and use.
\newblock {\em Rheol. Bull}, 79(2):14--18, 2010.

\bibitem{poole2012}
RJ~Poole.
\newblock The deborah and weissenberg numbers.
\newblock {\em British Soc. Rheol. Rheol. Bull}, 53:32--39, 2012.

\bibitem{hyun2011}
Kyu Hyun, Manfred Wilhelm, Christopher~O Klein, Kwang~Soo Cho, Jung~Gun Nam,
  Kyung~Hyun Ahn, Seung~Jong Lee, Randy~H Ewoldt, and Gareth~H McKinley.
\newblock A review of nonlinear oscillatory shear tests: Analysis and
  application of large amplitude oscillatory shear (laos).
\newblock {\em Progress in Polymer Science}, 36(12):1697--1753, 2011.

\bibitem{rogers2018}
Simon Rogers.
\newblock Large amplitude oscillatory shear: simple to describe, hard to
  interpret.
\newblock {\em Physics Today}, 71(7):34--40, 2018.

\bibitem{cheng2010}
L-S Cheng and R-Y Cao.
\newblock Constitutive model of viscous-elastic polymer solution in porous
  media.
\newblock {\em Petroleum Science and Technology}, 28(11):1170--1177, 2010.

\bibitem{hoagland1989}
David~A Hoagland and RK~Prud'Homme.
\newblock Hydrodynamic chromatography as a probe of polymer dynamics during
  flow through porous media.
\newblock {\em Macromolecules}, 22(2):775--781, 1989.

\bibitem{deGennes1976}
PG~De~Gennes.
\newblock Scaling theory of polymer adsorption.
\newblock {\em Journal de physique}, 37(12):1445--1452, 1976.

\bibitem{zitha2001}
Pacelli~LJ Zitha, Guy Chauveteau, and Liliane L{\'e}ger.
\newblock Unsteady-state flow of flexible polymers in porous media.
\newblock {\em Journal of colloid and interface science}, 234(2):269--283,
  2001.

\bibitem{babcock2003}
Hazen~P Babcock, Rodrigo~E Teixeira, Joe~S Hur, Eric~SG Shaqfeh, and Steven
  Chu.
\newblock Visualization of molecular fluctuations near the critical point of
  the coil- stretch transition in polymer elongation.
\newblock {\em Macromolecules}, 36(12):4544--4548, 2003.

\bibitem{schroeder2003}
Charles~M Schroeder, Hazen~P Babcock, Eric~SG Shaqfeh, and Steven Chu.
\newblock Observation of polymer conformation hysteresis in extensional flow.
\newblock {\em Science}, 301(5639):1515--1519, 2003.

\bibitem{obey1999}
Timothy~M Obey and PC~Griffiths.
\newblock Polymer adsorption: Fundamentals.
\newblock {\em COSMETIC SCIENCE AND TECHNOLOGY SERIES}, pages 51--72, 1999.

\bibitem{rubenstein2003}
M~Rubenstein and RH~Colby.
\newblock {\em Polymer physics: Oxford university press}.
\newblock Oxford, UK, 2003.

\bibitem{zitha1995}
P~Zitha, G~Chauveteau, A~Zaitoun, et~al.
\newblock Permeability\~{} dependent propagation of polyacrylamides under
  near-wellbore flow conditions.
\newblock In {\em SPE International Symposium on Oilfield Chemistry}. Society
  of Petroleum Engineers, 1995.

\bibitem{darwish2003}
Mohamed~IM Darwish, John~E McCray, Peter~K Currie, and Pacelli~LJ Zitha.
\newblock Polymer-enhanced dnapl flushing from low-permeability media: An
  experimental study.
\newblock {\em Groundwater Monitoring \& Remediation}, 23(2):92--101, 2003.

\bibitem{grattoni2004}
CA~Grattoni, PF~Luckham, XD~Jing, L~Norman, and Robert~W Zimmerman.
\newblock Polymers as relative permeability modifiers: adsorption and the
  dynamic formation of thick polyacrylamide layers.
\newblock {\em Journal of petroleum science and engineering}, 45(3-4):233--245,
  2004.

\bibitem{chauveteau1981}
G~Chauveteau and M~Moan.
\newblock The onset of dilatant behaviour in non-inertial flow of dilute
  polymer solutions through channels with varying cross-sections.
\newblock {\em Journal de Physique Lettres}, 42(10):201--204, 1981.

\bibitem{chauveteau1984}
G~Chauveteau, M~Moan, and A~Magueur.
\newblock Thickening behaviour of dilute polymer solutions in non-inertial
  elongational flows.
\newblock {\em Journal of non-newtonian fluid mechanics}, 16(3):315--327, 1984.

\bibitem{prabhakar2004}
R~Prabhakar, Papanasamoorthy Sunthar, and J~Ravi Prakash.
\newblock Exploring the universal dynamics of dilute polymer solutions in
  extensional flows.
\newblock {\em Physica A: Statistical Mechanics and its Applications},
  339(1-2):34--39, 2004.

\bibitem{gupta2000}
Rahul~K Gupta, DA~Nguyen, and T~Sridhar.
\newblock Extensional viscosity of dilute polystyrene solutions: Effect of
  concentration and molecular weight.
\newblock {\em Physics of Fluids}, 12(6):1296--1318, 2000.

\bibitem{smith1998}
Douglas~E Smith and Steven Chu.
\newblock Response of flexible polymers to a sudden elongational flow.
\newblock {\em Science}, 281(5381):1335--1340, 1998.

\bibitem{pilitsis1989}
Stergios Pilitsis and Antony~N Beris.
\newblock Calculations of steady-state viscoelastic flow in an undulating tube.
\newblock {\em Journal of Non-Newtonian Fluid Mechanics}, 31(3):231--287, 1989.

\bibitem{pilitsis1991}
Stergios Pilitsis and Antony~N Beris.
\newblock Viscoelastic flow in an undulating tube. part ii. effects of high
  elasticity, large amplitude of undulation and inertia.
\newblock {\em Journal of non-newtonian fluid mechanics}, 39(3):375--405, 1991.

\bibitem{talwar1995}
Kapil~K Talwar and Bamin Khomami.
\newblock Flow of viscoelastic fluids past periodic square arrays of cylinders:
  inertial and shear thinning viscosity and elasticity effects.
\newblock {\em Journal of Non-Newtonian Fluid Mechanics}, 57(2-3):177--202,
  1995.

\bibitem{rye1993}
Hays~S Rye, Jonathan~M Dabora, Mark~A Quesada, Richard~A Mathies, and
  Alexander~N Glazer.
\newblock Fluorometric assay using dimeric dyes for double-and single-stranded
  dna and rna with picogram sensitivity.
\newblock {\em Analytical biochemistry}, 208(1):144--150, 1993.

\bibitem{teclemariam2007}
Nerayo~P Teclemariam, Victor~A Beck, Eric~SG Shaqfeh, and Susan~J Muller.
\newblock Dynamics of dna polymers in post arrays: Comparison of single
  molecule experiments and simulations.
\newblock {\em Macromolecules}, 40(10):3848--3859, 2007.

\bibitem{hemminger2010}
Orin~L Hemminger, Pouyan~E Boukany, Shi-Qing Wang, and LJ~Lee.
\newblock Flow pattern and molecular visualization of dna solutions through a
  4: 1 planar micro-contraction.
\newblock {\em Journal of Non-Newtonian Fluid Mechanics},
  165(23-24):1613--1624, 2010.

\bibitem{perkins1997}
Thomas~T Perkins, Douglas~E Smith, and Steven Chu.
\newblock Single polymer dynamics in an elongational flow.
\newblock {\em Science}, 276(5321):2016--2021, 1997.

\bibitem{rems2016}
Lea Rems, Durgesh Kawale, L~James Lee, and Pouyan~E Boukany.
\newblock Flow of dna in micro/nanofluidics: From fundamentals to applications.
\newblock {\em Biomicrofluidics}, 10(4):043403, 2016.

\bibitem{schroeder2018}
Charles~M Schroeder.
\newblock Single polymer dynamics for molecular rheology.
\newblock {\em Journal of Rheology}, 62(1):371--403, 2018.

\bibitem{francois2008}
Nicolas Fran{\c{c}}ois, David Lasne, Yacine Amarouchene, Brahim Lounis, and
  Hamid Kellay.
\newblock Drag enhancement with polymers.
\newblock {\em Physical review letters}, 100(1):018302, 2008.

\bibitem{francois2009}
Nicolas Fran{\c{c}}ois, Yacine Amarouchene, Brahim Lounis, and Hamid Kellay.
\newblock Polymer conformations and hysteretic stresses in nonstationary flows
  of polymer solutions.
\newblock {\em EPL (Europhysics Letters)}, 86(3):34002, 2009.

\bibitem{deGennes1974}
PG~De~Gennes.
\newblock Coil-stretch transition of dilute flexible polymers under ultrahigh
  velocity gradients.
\newblock {\em The Journal of Chemical Physics}, 60(12):5030--5042, 1974.

\bibitem{fuller1980}
GG~Fuller and LG~Leal.
\newblock Flow birefringence of dilute polymer solutions in two-dimensional
  flows.
\newblock {\em Rheologica Acta}, 19(5):580--600, 1980.

\bibitem{hunkeler1996}
D~Hunkeler, TQ~Nguyen, and HH~Kausch.
\newblock Polymer solutions under elongational flow: 1. birefringence
  characterization of transient and stagnation point elongational flows.
\newblock {\em Polymer}, 37(19):4257--4269, 1996.

\bibitem{menasveta1991}
Malika~J Menasveta and David~A Hoagland.
\newblock Light scattering from dilute poly (styrene) solutions in uniaxial
  elongational flow.
\newblock {\em Macromolecules}, 24(11):3427--3433, 1991.

\bibitem{goff1994}
R~Blake Goff, Steven~F Karel, and Robert~K Prud’homme.
\newblock Electric birefringence of flexible polymers in high fields: Brownian
  dynamics simulation.
\newblock {\em The Journal of chemical physics}, 100(3):2289--2297, 1994.

\bibitem{gerashchenko2005}
S~Gerashchenko, C~Chevallard, and V~Steinberg.
\newblock Single-polymer dynamics: Coil-stretch transition in a random flow.
\newblock {\em EPL (Europhysics Letters)}, 71(2):221, 2005.

\bibitem{tanner1975}
RI~Tanner.
\newblock Stresses in dilute solutions of bead-nonlinear-spring macromolecules.
  iii. friction coefficient varying with dumbbell extension.
\newblock {\em Transactions of the Society of Rheology}, 19(4):557--582, 1975.

\bibitem{magda1988}
JJ~Magda, RG~Larson, and ME~Mackay.
\newblock Deformation-dependent hydrodynamic interaction in flows of dilute
  polymer solutions.
\newblock {\em The Journal of chemical physics}, 89(4):2504--2513, 1988.

\bibitem{rallison1988}
JM~Rallison and EJ~Hinch.
\newblock Do we understand the physics in the constitutive equation?
\newblock {\em Journal of Non-Newtonian Fluid Mechanics}, 29:37--55, 1988.

\bibitem{liu2007}
Yonggang Liu, Yonggun Jun, and Victor Steinberg.
\newblock Longest relaxation times of double-stranded and single-stranded dna.
\newblock {\em Macromolecules}, 40(6):2172--2176, 2007.

\bibitem{hsiao2017}
Kai-Wen Hsiao, Chandi Sasmal, J~Ravi~Prakash, and Charles~M Schroeder.
\newblock Direct observation of dna dynamics in semidilute solutions in
  extensional flow.
\newblock {\em Journal of Rheology}, 61(1):151--167, 2017.

\bibitem{hur2001}
Joe~S Hur, Eric~SG Shaqfeh, Hazen~P Babcock, Douglas~E Smith, and Steven Chu.
\newblock Dynamics of dilute and semidilute dna solutions in the start-up of
  shear flow.
\newblock {\em Journal of Rheology}, 45(2):421--450, 2001.

\bibitem{ho1977}
Teh~Chung Ho and Morton~M Denn.
\newblock Stability of plane poiseuille flow of a highly elastic liquid.
\newblock {\em Journal of Non-Newtonian Fluid Mechanics}, 3(2):179--195, 1977.

\bibitem{byars1994}
Jeffrey~A Byars, Alparslan {\"O}ztekin, Robert~A Brown, and Gareth~H Mckinley.
\newblock Spiral instabilities in the flow of highly elastic fluids between
  rotating parallel disks.
\newblock {\em Journal of Fluid Mechanics}, 271:173--218, 1994.

\bibitem{casanellas2016}
Laura Casanellas, Manuel~A Alves, Robert~J Poole, Sandra Lerouge, and Anke
  Lindner.
\newblock The stabilizing effect of shear thinning on the onset of purely
  elastic instabilities in serpentine microflows.
\newblock {\em Soft matter}, 12(29):6167--6175, 2016.

\bibitem{chakraborty2004}
Jyoti Chakraborty, Nishith Verma, and RP~Chhabra.
\newblock Wall effects in flow past a circular cylinder in a plane channel: a
  numerical study.
\newblock {\em Chemical Engineering and Processing: Process Intensification},
  43(12):1529--1537, 2004.

\bibitem{galindo2012}
Francisco~J Galindo-Rosales, Laura Campo-Dea{\~n}o, FT~Pinho, E~Van~Bokhorst,
  PJ~Hamersma, M{\'o}nica~SN Oliveira, and MA~Alves.
\newblock Microfluidic systems for the analysis of viscoelastic fluid flow
  phenomena in porous media.
\newblock {\em Microfluidics and nanofluidics}, 12(1-4):485--498, 2012.

\bibitem{galindo2014}
Francisco~J Galindo-Rosales, Laura Campo-Dea{\~n}o, Patr{\'\i}cia~C Sousa,
  Vera~M Ribeiro, M{\'o}nica~SN Oliveira, Manuel~A Alves, and Fernando~T Pinho.
\newblock Viscoelastic instabilities in micro-scale flows.
\newblock {\em Experimental Thermal and Fluid Science}, 59:128--139, 2014.

\bibitem{haward2016elastic}
Simon~J Haward, Gareth~H McKinley, and Amy~Q Shen.
\newblock Elastic instabilities in planar elongational flow of monodisperse
  polymer solutions.
\newblock {\em Scientific reports}, 6:33029, 2016.

\bibitem{hulsen2005}
Martien~A Hulsen, Raanan Fattal, and Raz Kupferman.
\newblock Flow of viscoelastic fluids past a cylinder at high weissenberg
  number: stabilized simulations using matrix logarithms.
\newblock {\em Journal of Non-Newtonian Fluid Mechanics}, 127(1):27--39, 2005.

\bibitem{larson1992instabilities}
Ronald~G Larson.
\newblock Instabilities in viscoelastic flows.
\newblock {\em Rheologica Acta}, 31(3):213--263, 1992.

\bibitem{muller1989}
Susan~J Muller, Ronald~G Larson, and Eric~SG Shaqfeh.
\newblock A purely elastic transition in taylor-couette flow.
\newblock {\em Rheologica Acta}, 28(6):499--503, 1989.

\bibitem{larson1990}
Ronald~G Larson, Eric~SG Shaqfeh, and Susan~J Muller.
\newblock A purely elastic instability in taylor--couette flow.
\newblock {\em Journal of Fluid Mechanics}, 218:573--600, 1990.

\bibitem{lee1991}
T~Lee and R~Budwig.
\newblock A study of the effect of aspect ratio on vortex shedding behind
  circular cylinders.
\newblock {\em Physics of Fluids A: Fluid Dynamics}, 3(2):309--315, 1991.

\bibitem{li2012}
Xiao-Bin Li, Feng-Chen Li, Wei-Hua Cai, Hong-Na Zhang, and Juan-Cheng Yang.
\newblock Very-low-re chaotic motions of viscoelastic fluid and its unique
  applications in microfluidic devices: a review.
\newblock {\em Experimental thermal and fluid science}, 39:1--16, 2012.

\bibitem{magda1988transition}
JJ~Magda and RG~Larson.
\newblock A transition occurring in ideal elastic liquids during shear flow.
\newblock {\em Journal of non-newtonian fluid mechanics}, 30(1):1--19, 1988.

\bibitem{mckinley1993}
Gareth~H McKinley, Robert~C Armstrong, and Robert Brown.
\newblock The wake instability in viscoelastic flow past confined circular
  cylinders.
\newblock {\em Philosophical Transactions of the Royal Society of London.
  Series A: Physical and Engineering Sciences}, 344(1671):265--304, 1993.

\bibitem{mckinley1991}
Gareth~H McKinley, Jeffrey~A Byars, Robert~A Brown, and Robert~C Armstrong.
\newblock Observations on the elastic instability in cone-and-plate and
  parallel-plate flows of a polyisobutylene boger fluid.
\newblock {\em Journal of Non-Newtonian Fluid Mechanics}, 40(2):201--229, 1991.

\bibitem{mckinley1995}
Gareth~H Mckinley, Alparslan {\"O}ztekin, Jeffrey~A Byars, and Robert~A Brown.
\newblock Self-similar spiral instabilities in elastic flows between a cone and
  a plate.
\newblock {\em Journal of Fluid Mechanics}, 285:123--164, 1995.

\bibitem{oliveira2005}
Paulo~J Oliveira and Am{\'\i}lcar~IP Miranda.
\newblock A numerical study of steady and unsteady viscoelastic flow past
  bounded cylinders.
\newblock {\em Journal of non-newtonian fluid mechanics}, 127(1):51--66, 2005.

\bibitem{oliveira2009}
MSN Oliveira, FT~Pinho, RJ~Poole, PJ~Oliveira, and MA~Alves.
\newblock Purely elastic flow asymmetries in flow-focusing devices.
\newblock {\em Journal of Non-Newtonian Fluid Mechanics}, 160(1):31--39, 2009.

\bibitem{poole2007}
RJ~Poole, MA~Alves, and Paulo~J Oliveira.
\newblock Purely elastic flow asymmetries.
\newblock {\em Physical review letters}, 99(16):164503, 2007.

\bibitem{pathak2004}
Jai~A Pathak, David Ross, and Kalman~B Migler.
\newblock Elastic flow instability, curved streamlines, and mixing in
  microfluidic flows.
\newblock {\em Physics of fluids}, 16(11):4028--4034, 2004.

\bibitem{ribeiro2014}
VM~Ribeiro, PM~Coelho, FT~Pinho, and MA~Alves.
\newblock Viscoelastic fluid flow past a confined cylinder: Three-dimensional
  effects and stability.
\newblock {\em Chemical engineering science}, 111:364--380, 2014.

\bibitem{sen2009}
Subhankar Sen, Sanjay Mittal, and Gautam Biswas.
\newblock Steady separated flow past a circular cylinder at low reynolds
  numbers.
\newblock {\em Journal of Fluid Mechanics}, 620:89--119, 2009.

\bibitem{shaqfeh1996}
Eric~SG Shaqfeh.
\newblock Purely elastic instabilities in viscometric flows.
\newblock {\em Annual Review of Fluid Mechanics}, 28(1):129--185, 1996.

\bibitem{williamson1996}
Charles~HK Williamson.
\newblock Vortex dynamics in the cylinder wake.
\newblock {\em Annual review of fluid mechanics}, 28(1):477--539, 1996.

\bibitem{wilson1999}
Helen~J Wilson, Michael Renardy, and Yuriko Renardy.
\newblock Structure of the spectrum in zero reynolds number shear flow of the
  ucm and oldroyd-b liquids.
\newblock {\em Journal of non-newtonian fluid mechanics}, 80(2-3):251--268,
  1999.

\bibitem{chilcott1988}
MD~Chilcott and JM~Rallison.
\newblock Creeping flow of dilute polymer solutions past cylinders and spheres.
\newblock {\em Journal of non-newtonian fluid mechanics}, 29:381--432, 1988.

\bibitem{haward2019}
Simon~J Haward, Naoyuki Kitajima, Kazumi Toda-Peters, Tsutomu Takahashi, and
  Amy~Q Shen.
\newblock Flow of wormlike micellar solutions around microfluidic cylinders
  with high aspect ratio and low blockage ratio.
\newblock {\em Soft Matter}, 15:1927--1941, 2019.

\bibitem{qin2017}
Boyang Qin and Paulo~E Arratia.
\newblock Characterizing elastic turbulence in channel flows at low reynolds
  number.
\newblock {\em Physical Review Fluids}, 2(8):083302, 2017.

\bibitem{berti2008}
S~Berti, A~Bistagnino, Guido Boffetta, A~Celani, and S~Musacchio.
\newblock Two-dimensional elastic turbulence.
\newblock {\em Physical Review E}, 77(5):055306, 2008.

\bibitem{grilli2013}
Muzio Grilli, Adolfo V{\'a}zquez-Quesada, and Marco Ellero.
\newblock Transition to turbulence and mixing in a viscoelastic fluid flowing
  inside a channel with a periodic array of cylindrical obstacles.
\newblock {\em Physical review letters}, 110(17):174501, 2013.

\bibitem{zami2017}
Fr{\'e}d{\'e}ric Zami-Pierre, R~de~Loubens, Michel Quintard, and Yohan Davit.
\newblock Polymer flow through porous media: numerical prediction of the
  contribution of slip to the apparent viscosity.
\newblock {\em Transport in Porous Media}, 119(3):521--538, 2017.

\bibitem{boffetta2003}
Guido Boffetta, Antonio Celani, and Stefano Musacchio.
\newblock Two-dimensional turbulence of dilute polymer solutions.
\newblock {\em Physical review letters}, 91(3):034501, 2003.

\bibitem{liu2014}
Yonggang Liu and Victor Steinberg.
\newblock Single polymer dynamics in a random flow.
\newblock In {\em Macromolecular Symposia}, volume 337, pages 34--43. Wiley
  Online Library, 2014.

\bibitem{haward2013}
SJ~Haward and GH~McKinley.
\newblock Instabilities in stagnation point flows of polymer solutions.
\newblock {\em Physics of Fluids}, 25(8):083104, 2013.

\bibitem{sousa2015}
PC~Sousa, FT~Pinho, MSN Oliveira, and MA~Alves.
\newblock Purely elastic flow instabilities in microscale cross-slot devices.
\newblock {\em Soft matter}, 11(45):8856--8862, 2015.

\bibitem{lanzaro2017}
Alfredo Lanzaro, Daniel Corbett, and Xue-Feng Yuan.
\newblock Non-linear dynamics of semi-dilute paam solutions in a microfluidic
  3d cross-slot flow geometry.
\newblock {\em Journal of Non-Newtonian Fluid Mechanics}, 242:57--65, 2017.

\bibitem{jovanovic2011}
Mihailo~R Jovanovi{\'c} and Satish Kumar.
\newblock Nonmodal amplification of stochastic disturbances in strongly elastic
  channel flows.
\newblock {\em Journal of Non-Newtonian Fluid Mechanics}, 166(14-15):755--778,
  2011.

\bibitem{arratia2006}
Paulo~E Arratia, CC~Thomas, J~Diorio, and Jerry~P Gollub.
\newblock Elastic instabilities of polymer solutions in cross-channel flow.
\newblock {\em Physical review letters}, 96(14):144502, 2006.

\bibitem{pan2013}
L~Pan, A~Morozov, C~Wagner, and PE~Arratia.
\newblock Nonlinear elastic instability in channel flows at low reynolds
  numbers.
\newblock {\em Physical review letters}, 110(17):174502, 2013.

\bibitem{pakdel1996}
Peyman Pakdel and Gareth~H McKinley.
\newblock Elastic instability and curved streamlines.
\newblock {\em Physical Review Letters}, 77(12):2459, 1996.

\bibitem{qin2018}
Boyang Qin, Paul~F Salipante, Steven~D Hudson, and Paulo~E Arratia.
\newblock Flow resistance and structures in viscoelastic channel flows at low
  re.
\newblock {\em arXiv preprint arXiv:1807.00927}, 2018.

\bibitem{mckinley1996}
Gareth~H McKinley, Peyman Pakdel, and Alparslan {\"O}ztekin.
\newblock Rheological and geometric scaling of purely elastic flow
  instabilities.
\newblock {\em Journal of Non-Newtonian Fluid Mechanics}, 67:19--47, 1996.

\bibitem{zilz2012}
J~Zilz, RJ~Poole, MA~Alves, D~Bartolo, B~Levach{\'e}, and A~Lindner.
\newblock Geometric scaling of a purely elastic flow instability in serpentine
  channels.
\newblock {\em Journal of Fluid Mechanics}, 712:203--218, 2012.

\bibitem{lanzaro2011}
Alfredo Lanzaro and Xue-Feng Yuan.
\newblock Effects of contraction ratio on non-linear dynamics of semi-dilute,
  highly polydisperse paam solutions in microfluidics.
\newblock {\em Journal of Non-Newtonian Fluid Mechanics},
  166(17-18):1064--1075, 2011.

\bibitem{rodd2005}
Lucy~E Rodd, Timothy~P Scott, David~V Boger, Justin~J Cooper-White, and
  Gareth~H McKinley.
\newblock The inertio-elastic planar entry flow of low-viscosity elastic fluids
  in micro-fabricated geometries.
\newblock {\em Journal of Non-Newtonian Fluid Mechanics}, 129(1):1--22, 2005.

\bibitem{koelling1991}
KW~Koelling and Robert~Krafft Prud'homme.
\newblock Instabilities in multi-hole converging flow of viscoelastic fluids.
\newblock {\em Rheologica acta}, 30(6):511--522, 1991.

\bibitem{lanzaro2011thesis}
Alfredo Lanzaro.
\newblock {\em Microscopic flows of aqueous polyacrylamide solutions: a
  quantitative study}.
\newblock PhD thesis, The University of Manchester (United Kingdom), 2011.

\bibitem{lanzaro2014}
Alfredo Lanzaro and Xue-Feng Yuan.
\newblock A quantitative analysis of spatial extensional rate distribution in
  nonlinear viscoelastic flows.
\newblock {\em Journal of Non-Newtonian Fluid Mechanics}, 207:32--41, 2014.

\bibitem{lanzaro2015}
Alfredo Lanzaro, Zhuo Li, and Xue-Feng Yuan.
\newblock Quantitative characterization of high molecular weight polymer
  solutions in microfluidic hyperbolic contraction flow.
\newblock {\em Microfluidics and Nanofluidics}, 18(5-6):819--828, 2015.

\bibitem{gulati2015}
Shelly Gulati, Susan~J Muller, and Dorian Liepmann.
\newblock Flow of dna solutions in a microfluidic gradual contraction.
\newblock {\em Biomicrofluidics}, 9(5):054102, 2015.

\bibitem{szabo1997}
Peter Szabo, JM~Rallison, and EJ~Hinch.
\newblock Start-up of flow of a fene-fluid through a 4: 1: 4 constriction in a
  tube.
\newblock {\em Journal of non-newtonian fluid mechanics}, 72(1):73--86, 1997.

\bibitem{qin2019}
Boyang Qin, Paul~F Salipante, Steven~D Hudson, and Paulo~E Arratia.
\newblock Upstream vortex and elastic wave in the viscoelastic flow around a
  confined cylinder.
\newblock {\em Journal of Fluid Mechanics}, 864, 2019.

\bibitem{varshney2017}
Atul Varshney and Victor Steinberg.
\newblock Elastic wake instabilities in a creeping flow between two obstacles.
\newblock {\em Physical Review Fluids}, 2(5):051301, 2017.

\bibitem{haward2018}
Simon~J Haward, Kazumi Toda-Peters, and Amy~Q Shen.
\newblock Steady viscoelastic flow around high-aspect-ratio, low-blockage-ratio
  microfluidic cylinders.
\newblock {\em Journal of Non-Newtonian Fluid Mechanics}, 254:23--35, 2018.

\bibitem{khomami1997}
Barnin Khomami and Luis~D Moreno.
\newblock Stability of viscoelastic flow around periodic arrays of cylinders.
\newblock {\em Rheologica acta}, 36(4):367--383, 1997.

\bibitem{arora2002}
K~Arora, R~Sureshkumar, and B~Khomami.
\newblock Experimental investigation of purely elastic instabilities in
  periodic flows.
\newblock {\em Journal of non-newtonian fluid mechanics}, 108(1-3):209--226,
  2002.

\bibitem{shi2015}
Xueda Shi, Stephen Kenney, Ganesh Chapagain, and Gordon~F Christopher.
\newblock Mechanisms of onset for moderate mach number instabilities of
  viscoelastic flows around confined cylinders.
\newblock {\em Rheologica Acta}, 54(9-10):805--815, 2015.

\bibitem{shi2016}
Xueda Shi and Gordon~F Christopher.
\newblock Growth of viscoelastic instabilities around linear cylinder arrays.
\newblock {\em Physics of Fluids}, 28(12):124102, 2016.

\bibitem{kawale2019}
Durgesh Kawale, Jishnu Jayaraman, and Pouyan~E Boukany.
\newblock Microfluidic rectifier for polymer solutions flowing through porous
  media.
\newblock {\em Biomicrofluidics}, 13(1):014111, 2019.

\bibitem{de2016}
S~De, J~van~der Schaaf, NG~Deen, JAM Kuipers, EAJF Peters, and JT~Padding.
\newblock Lane change in flows through pillared microchannels.
\newblock {\em Physics of Fluids}, 29:113102 1--8, 2017.

\bibitem{de2018viscoelastic}
S~De, P~Krishnan, J~van~der Schaaf, JAM Kuipers, EAJF Peters, and JT~Padding.
\newblock Viscoelastic effects on residual oil distribution in flows through
  pillared microchannels.
\newblock {\em Journal of colloid and interface science}, 510:262--271, 2018.

\bibitem{sadanandan2004}
B~Sadanandan and Radhakrishna Sureshkumar.
\newblock Global linear stability analysis of viscoelastic flow through a
  periodic channel.
\newblock {\em Journal of non-newtonian fluid mechanics}, 122(1-3):55--67,
  2004.

\bibitem{de2016coupled}
S~De, S~Das, JAM Kuipers, EAJF Peters, and JT~Padding.
\newblock A coupled finite volume immersed boundary method for simulating 3d
  viscoelastic flows in complex geometries.
\newblock {\em Journal of Non-Newtonian Fluid Mechanics}, 232:67--76, 2016.

\bibitem{de2017b}
S~De, JAM Kuipers, EAJF Peters, and JT~Padding.
\newblock Viscoelastic flow simulations in model porous media.
\newblock {\em Physical Review Fluids}, 2(5):053303, 2017.

\bibitem{vazquez2012}
A~V{\'a}zquez-Quesada and M~Ellero.
\newblock Sph simulations of a viscoelastic flow around a periodic array of
  cylinders confined in a channel.
\newblock {\em Journal of Non-Newtonian Fluid Mechanics}, 167:1--8, 2012.

\bibitem{de2017a}
S~De, JAM Kuipers, EAJF Peters, and JT~Padding.
\newblock Viscoelastic flow simulations in random porous media.
\newblock {\em Journal of Non-Newtonian Fluid Mechanics}, 248:50--61, 2017.

\bibitem{de2017viscoelastic}
Shauvik De, Johannes~AM Kuipers, Elias~AJF Peters, and Johan~T Padding.
\newblock Viscoelastic flow past mono-and bidisperse random arrays of
  cylinders: flow resistance, topology and normal stress distribution.
\newblock {\em Soft matter}, 13(48):9138--9146, 2017.

\bibitem{walkama2019}
Derek~M Walkama, Nicolas Waisbord, and Jeffrey~S Guasto.
\newblock Disorder suppresses chaos in viscoelastic flows.
\newblock {\em arXiv preprint arXiv:1906.11868}, 2019.

\bibitem{datta2014b}
Sujit~S Datta, TS~Ramakrishnan, and David~A Weitz.
\newblock Mobilization of a trapped non-wetting fluid from a three-dimensional
  porous medium.
\newblock {\em Physics of Fluids}, 26(2):022002, 2014.

\bibitem{morrow1988}
NR~Morrow, I~Chatzis, JJ~Taber, et~al.
\newblock Entrapment and mobilization of residual oil in bead packs.
\newblock {\em SPE Reservoir Engineering}, 3(03):927--934, 1988.

\bibitem{amili2006}
P~Amili and YC~Yortsos.
\newblock Darcian dynamics: a new approach to the mobilization of a trapped
  phase in porous media.
\newblock {\em Transport in porous media}, 64(1):25--49, 2006.

\bibitem{wang2010}
Demin Wang, Huifen Xia, Shuren Yang, Gang Wang, et~al.
\newblock The influence of visco-elasticity on micro forces and displacement
  efficiency in pores, cores and in the field.
\newblock In {\em SPE EOR Conference at Oil \& Gas West Asia}. Society of
  Petroleum Engineers, 2010.

\bibitem{giese2002}
Steven~W Giese and Susan~E Powers.
\newblock Using polymer solutions to enhance recovery of mobile coal tar and
  creosote dnapls.
\newblock {\em Journal of contaminant hydrology}, 58(1-2):147--167, 2002.

\bibitem{smith2008}
Megan~M Smith, Jeff~AK Silva, Junko Munakata-Marr, and John~E McCray.
\newblock Compatibility of polymers and chemical oxidants for enhanced
  groundwater remediation.
\newblock {\em Environmental science \& technology}, 42(24):9296--9301, 2008.

\bibitem{durst1981}
FRBU Durst, R~Haas, and BU~Kaczmar.
\newblock Flows of dilute hydrolyzed polyacrylamide solutions in porous media
  under various solvent conditions.
\newblock {\em Journal of Applied Polymer Science}, 26(9):3125--3149, 1981.

\bibitem{schneider1982}
FN~Schneider, WW~Owens, et~al.
\newblock Steady-state measurements of relative permeability for polymer/oil
  systems.
\newblock {\em Society of Petroleum Engineers Journal}, 22(01):79--86, 1982.

\bibitem{sandengen2017}
K~Sandengen, MT~Tweheyo, CM~Crescente, A~Mouret, I~Henaut, and D~Rousseau.
\newblock Qualifying an “emulsion” polymer for field use-lab-scale
  assessments on adsorption and injectivity.
\newblock In {\em IOR 2017-19th European Symposium on Improved Oil Recovery},
  2017.

\bibitem{xu2017}
Ke~Xu, Tianbo Liang, Peixi Zhu, Pengpeng Qi, Jun Lu, Chun Huh, and Matthew
  Balhoff.
\newblock A 2.5-d glass micromodel for investigation of multi-phase flow in
  porous media.
\newblock {\em Lab on a Chip}, 17(4):640--646, 2017.

\bibitem{lacey2017}
Mike Lacey, Cathy Hollis, Mart Oostrom, and Nima Shokri.
\newblock Effects of pore and grain size on water and polymer flooding in
  micromodels.
\newblock {\em Energy \& Fuels}, 31(9):9026--9034, 2017.

\bibitem{avendano2013}
J~Avendano, Nicolas Pannacci, Benjamin Herzhaft, Patrick Gateau, and P~Coussot.
\newblock Enhanced displacement of a liquid pushed by a viscoelastic fluid.
\newblock {\em Journal of colloid and interface science}, 410:172--180, 2013.

\bibitem{roof1970}
JG~Roof et~al.
\newblock Snap-off of oil droplets in water-wet pores.
\newblock {\em Society of Petroleum Engineers Journal}, 10(01):85--90, 1970.

\bibitem{derzsi2013}
Ladislav Derzsi, Marta Kasprzyk, Jan~Philip Plog, and Piotr Garstecki.
\newblock Flow focusing with viscoelastic liquids.
\newblock {\em Physics of Fluids}, 25(9):092001, 2013.

\bibitem{gupta2016}
Anupam Gupta and Mauro Sbragaglia.
\newblock Effects of viscoelasticity on droplet dynamics and break-up in
  microfluidic t-junctions: a lattice boltzmann study.
\newblock {\em The European Physical Journal E}, 39(1):6, 2016.

\bibitem{nooranidoost2016}
Mohammad Nooranidoost, Daulet Izbassarov, and Metin Muradoglu.
\newblock Droplet formation in a flow focusing configuration: Effects of
  viscoelasticity.
\newblock {\em Physics of Fluids}, 28(12):123102, 2016.

\bibitem{nekouei2017}
Mehdi Nekouei and Siva~A Vanapalli.
\newblock Volume-of-fluid simulations in microfluidic t-junction devices:
  Influence of viscosity ratio on droplet size.
\newblock {\em Physics of Fluids}, 29(3):032007, 2017.

\bibitem{clarke2015}
Andrew Clarke, Andrew~M Howe, Jonathan Mitchell, John Staniland, Laurence
  Hawkes, and Katherine Leeper.
\newblock Mechanism of anomalously increased oil displacement with aqueous
  viscoelastic polymer solutions.
\newblock {\em Soft Matter}, 11(18):3536--3541, 2015.

\bibitem{de2018flow}
S~De, SP~Koesen, RV~Maitri, M~Golombok, JT~Padding, and JFM van Santvoort.
\newblock Flow of viscoelastic surfactants through porous media.
\newblock {\em AIChE Journal}, 64(2):773--781, 2018.

\bibitem{barnes1995}
Howard~A Barnes.
\newblock A review of the slip (wall depletion) of polymer solutions, emulsions
  and particle suspensions in viscometers: its cause, character, and cure.
\newblock {\em Journal of Non-Newtonian Fluid Mechanics}, 56(3):221--251, 1995.

\bibitem{sochi2011}
Taha Sochi.
\newblock Slip at fluid-solid interface.
\newblock {\em Polymer Reviews}, 51(4):309--340, 2011.

\bibitem{fang2005}
Lin Fang, Hua Hu, and Ronald~G Larson.
\newblock Dna configurations and concentration in shearing flow near a glass
  surface in a microchannel.
\newblock {\em Journal of rheology}, 49(1):127--138, 2005.

\bibitem{ma2005}
Hongbo Ma and Michael~D Graham.
\newblock Theory of shear-induced migration in dilute polymer solutions near
  solid boundaries.
\newblock {\em Physics of Fluids}, 17(8):083103, 2005.

\bibitem{kekre2010}
Rahul Kekre, Jason~E Butler, and Anthony~JC Ladd.
\newblock Comparison of lattice-boltzmann and brownian-dynamics simulations of
  polymer migration in confined flows.
\newblock {\em Physical Review E}, 82(1):011802, 2010.

\bibitem{joanny1979}
JF~Joanny, L~Leibler, and PG~De~Gennes.
\newblock Effects of polymer solutions on colloid stability.
\newblock {\em Journal of Polymer Science: Polymer Physics Edition},
  17(6):1073--1084, 1979.

\bibitem{uematsu2015}
Yuki Uematsu.
\newblock Nonlinear electro-osmosis of dilute non-adsorbing polymer solutions
  with low ionic strength.
\newblock {\em Soft matter}, 11(37):7402--7411, 2015.

\bibitem{zhang2010}
Changyong Zhang, Karl Dehoff, Nancy Hess, Mart Oostrom, Thomas~W Wietsma,
  Albert~J Valocchi, Bruce~W Fouke, and Charles~J Werth.
\newblock Pore-scale study of transverse mixing induced caco3 precipitation and
  permeability reduction in a model subsurface sedimentary system.
\newblock {\em Environmental science \& technology}, 44(20):7833--7838, 2010.

\bibitem{song2014}
Wen Song, Thomas~W de~Haas, Hossein Fadaei, and David Sinton.
\newblock Chip-off-the-old-rock: the study of reservoir-relevant geological
  processes with real-rock micromodels.
\newblock {\em Lab on a Chip}, 14(22):4382--4390, 2014.

\bibitem{singh2017}
Rajveer Singh, Mayandi Sivaguru, Glenn~A Fried, Bruce~W Fouke, Robert~A
  Sanford, Martin Carrera, and Charles~J Werth.
\newblock Real rock-microfluidic flow cell: A test bed for real-time in situ
  analysis of flow, transport, and reaction in a subsurface reactive transport
  environment.
\newblock {\em Journal of contaminant hydrology}, 204:28--39, 2017.

\bibitem{gerold2018}
Chase~T Gerold, Amber~T Krummel, and Charles~S Henry.
\newblock Microfluidic devices containing thin rock sections for oil recovery
  studies.
\newblock {\em Microfluidics and Nanofluidics}, 22(7):76, 2018.

\bibitem{datta2013a}
Sujit~S Datta, Harry Chiang, TS~Ramakrishnan, and David~A Weitz.
\newblock Spatial fluctuations of fluid velocities in flow through a
  three-dimensional porous medium.
\newblock {\em Physical review letters}, 111(6):064501, 2013.

\bibitem{krummel2013}
Amber~T Krummel, Sujit~S Datta, Stefan M{\"u}nster, and David~A Weitz.
\newblock Visualizing multiphase flow and trapped fluid configurations in a
  model three-dimensional porous medium.
\newblock {\em AIChE Journal}, 59(3):1022--1029, 2013.

\bibitem{datta2014a}
Sujit~S Datta, Jean-Baptiste Dupin, and David~A Weitz.
\newblock Fluid breakup during simultaneous two-phase flow through a
  three-dimensional porous medium.
\newblock {\em Physics of Fluids}, 26(6):062004, 2014.

\bibitem{groisman2001}
Alexander Groisman and Victor Steinberg.
\newblock Efficient mixing at low reynolds numbers using polymer additives.
\newblock {\em Nature}, 410(6831):905, 2001.

\bibitem{groisman2004}
Alexander Groisman and Victor Steinberg.
\newblock Elastic turbulence in curvilinear flows of polymer solutions.
\newblock {\em New Journal of Physics}, 6(1):29, 2004.

\bibitem{scholz2014}
Christian Scholz, Frank Wirner, Juan~Ruben Gomez-Solano, and Clemens Bechinger.
\newblock Enhanced dispersion by elastic turbulence in porous media.
\newblock {\em EPL (Europhysics Letters)}, 107(5):54003, 2014.

\bibitem{babayekhorasani2016}
Firoozeh Babayekhorasani, Dave~E Dunstan, Ramanan Krishnamoorti, and Jacinta~C
  Conrad.
\newblock Nanoparticle dispersion in disordered porous media with and without
  polymer additives.
\newblock {\em Soft Matter}, 12(26):5676--5683, 2016.

\bibitem{aramideh2019}
Soroush Aramideh, Pavlos~P Vlachos, and Arezoo~M Ardekani.
\newblock Nanoparticle dispersion in porous media in viscoelastic polymer
  solutions.
\newblock {\em Journal of Non-Newtonian Fluid Mechanics}, 268:75--80, 2019.

\bibitem{burghelea2004}
Teodor Burghelea, Enrico Segre, Israel Bar-Joseph, Alex Groisman, and Victor
  Steinberg.
\newblock Chaotic flow and efficient mixing in a microchannel with a polymer
  solution.
\newblock {\em Physical Review E}, 69(6):066305, 2004.

\bibitem{lam2009}
YC~Lam, HY~Gan, Nam-Trung Nguyen, and H~Lie.
\newblock Micromixer based on viscoelastic flow instability at low reynolds
  number.
\newblock {\em Biomicrofluidics}, 3(1):014106, 2009.

\bibitem{ligrani2017}
Phil Ligrani, Daniel Copeland, Chong Ren, Mengying Su, and Masaaki Suzuki.
\newblock Heat transfer enhancements from elastic turbulence using
  sucrose-based polymer solutions.
\newblock {\em Journal of Thermophysics and Heat Transfer}, 32(1):51--60, 2017.

\bibitem{bourgeat2003}
Alain Bourgeat, Olivier Gipouloux, and Eduard Marusic-Paloka.
\newblock Filtration law for polymer flow through porous media.
\newblock {\em Multiscale Modeling \& Simulation}, 1(3):432--457, 2003.

\bibitem{diao2014}
Ying Diao, Leo Shaw, Zhenan Bao, and Stefan~CB Mannsfeld.
\newblock Morphology control strategies for solution-processed organic
  semiconductor thin films.
\newblock {\em Energy \& Environmental Science}, 7(7):2145--2159, 2014.

\bibitem{compton2014}
Brett~G Compton and Jennifer~A Lewis.
\newblock 3d-printing of lightweight cellular composites.
\newblock {\em Advanced materials}, 26(34):5930--5935, 2014.

\bibitem{wang20173d}
Xin Wang, Man Jiang, Zuowan Zhou, Jihua Gou, and David Hui.
\newblock 3d printing of polymer matrix composites: A review and prospective.
\newblock {\em Composites Part B: Engineering}, 110:442--458, 2017.

\bibitem{luo1996}
Min Luo and Iwao Teraoka.
\newblock High osmotic pressure chromatography for large-scale fractionation of
  polymers.
\newblock {\em Macromolecules}, 29(12):4226--4233, 1996.

\bibitem{vanapalli2006}
Siva~A Vanapalli, Steven~L Ceccio, and Michael~J Solomon.
\newblock Universal scaling for polymer chain scission in turbulence.
\newblock {\em Proceedings of the National Academy of Sciences},
  103(45):16660--16665, 2006.

\end{thebibliography}
\bibliographystyle{unsrt} 

\end{document}